\documentclass[%
 reprint,
superscriptaddress,
 amsmath,amssymb,
 aps,
prb,
longbibliography,
]{revtex4-1}

\usepackage[colorlinks=true,linkcolor=blue,urlcolor=blue,citecolor=blue]{hyperref}
\usepackage{graphicx}
\usepackage{dcolumn}
\usepackage{bm}
\usepackage{multirow}
\usepackage{float}
\usepackage{setspace}
\usepackage{longtable}
\usepackage{booktabs}

\newcommand\degree{^\circ}

\begin{document}


\title{Varying magnetism in the lattice distorted Y$_{2}$NiIrO$_{6}$ and La$_{2}$NiIrO$_{6}$}


\author{Lu Liu}
\thanks{L. L. and K. Y. contributed equally to this paper.}
 \affiliation{Laboratory for Computational Physical Sciences (MOE),
 State Key Laboratory of Surface Physics, and Department of Physics,
  Fudan University, Shanghai 200433, China}
\affiliation{Shanghai Qi Zhi Institute, Shanghai 200232, China}

\author{Ke Yang}
\thanks{L. L. and K. Y. contributed equally to this paper.}
\affiliation{College of Science, University of Shanghai for Science and Technology, Shanghai 200093, China}
 \affiliation{Laboratory for Computational Physical Sciences (MOE),
 State Key Laboratory of Surface Physics, and Department of Physics,
  Fudan University, Shanghai 200433, China}

\author{Di Lu}
 \affiliation{Laboratory for Computational Physical Sciences (MOE),
 State Key Laboratory of Surface Physics, and Department of Physics,
  Fudan University, Shanghai 200433, China}
\affiliation{Shanghai Qi Zhi Institute, Shanghai 200232, China}

\author{Yaozhenghang Ma}
 \affiliation{Laboratory for Computational Physical Sciences (MOE),
 State Key Laboratory of Surface Physics, and Department of Physics,
  Fudan University, Shanghai 200433, China}
\affiliation{Shanghai Qi Zhi Institute, Shanghai 200232, China}

\author{Yuxuan Zhou}
 \affiliation{Laboratory for Computational Physical Sciences (MOE),
 State Key Laboratory of Surface Physics, and Department of Physics,
  Fudan University, Shanghai 200433, China}
\affiliation{Shanghai Qi Zhi Institute, Shanghai 200232, China}

\author{Hua Wu}
\thanks{Corresponding author: wuh@fudan.edu.cn}
\affiliation{Laboratory for Computational Physical Sciences (MOE),
 State Key Laboratory of Surface Physics, and Department of Physics,
 Fudan University, Shanghai 200433, China}
 \affiliation{Shanghai Qi Zhi Institute, Shanghai 200232, China}
\affiliation{Collaborative Innovation Center of Advanced Microstructures,
 Nanjing 210093, China}

\date{\today}

\begin{abstract}
We investigate the electronic and magnetic properties of the newly synthesized double perovskites Y$_{2}$NiIrO$_{6}$ and La$_{2}$NiIrO$_{6}$, using density functional calculations, crystal field theory, superexchange pictures, and Monte Carlo simulations. We find that both systems are antiferromagnetic (AFM) Mott insulators, with the high-spin Ni$^{2+}$ $t_{2g}$$^{6}e_{g}$$^{2}$ ($S=1$) and the low-spin Ir$^{4+}$ $t_{2g}$$^{5}$ ($S=1/2$) configurations. We address that their lattice distortion induces $t_{2g}$-$e_{g}$ orbital mixing and thus enables the normal Ni$^{+}$-Ir$^{5+}$ charge excitation with the electron hopping from the Ir `$t_{2g}$' to Ni `$e_g$' orbitals, which promotes the AFM Ni$^{2+}$-Ir$^{4+}$ coupling. Therefore, the increasing $t_{2g}$-$e_{g}$ mixing accounts for the enhanced $T_{\rm N}$ from the less distorted La$_{2}$NiIrO$_{6}$ to the more distorted Y$_{2}$NiIrO$_{6}$. Moreover, our test calculations find that in the otherwise ideally cubic Y$_{2}$NiIrO$_{6}$, the Ni$^{+}$-Ir$^{5+}$ charge excitation is forbidden, and only the abnormal Ni$^{3+}$-Ir$^{3+}$ excitation gives a weakly ferromagnetic (FM) behavior. Furthermore, we find that owing to the crystal field splitting, Hund exchange, and broad band formation in the highly coordinated fcc sublattice, Ir$^{4+}$ ions are not in the $j_{\rm eff}=1/2$ state but in the $S=1/2$ state carrying a finite orbital moment by spin-orbit coupling (SOC). This paper clarifies the varying magnetism in Y$_{2}$NiIrO$_{6}$ and La$_{2}$NiIrO$_{6}$ associated with the lattice distortions. 
\end{abstract}

\maketitle

\section{Introduction}
Perovskite oxides of the form $AB$O$_{3}$ involving a transition metal (TM) at the $B$ site often possess charge, spin, orbital, and lattice degrees of freedom\cite{Tokura_2000Science}. The interplay of these degrees of freedom provides a great platform for exploring intriguing properties and exotic phases, such as colossal magnetoresistance, multiferroicity, and superconductivity\cite{mr,ferroelectric,multiferroics,superc}. In recent decades, double perovskites $A_{2}BB'$O$_{6}$ have further enriched the magnetic and electronic properties of perovskite oxides by enabling varying combinations of two different TM atoms\cite{Vasala_2015}. In particular, the hybrid $3d$-$5d$ TM double perovskites combine the possibly strong correlation effect of $3d$ electrons with the pronounced SOC effect of $5d$ electrons, leading to novel properties that attract a wide range of interests. For instance, Sr$_{2}$CrReO$_{6}$ is a FM half-metal with high Curie temperature $T_{\rm C}$ of 635 K\cite{Sr2CrReO6}; Sr$_{2}$CrOsO$_{6}$ is a ferromagnetic (FM) insulator with $T_{\rm C}$ of 725 K\cite{Sr2CrOsO6}. Upon substituting Sr$^{2+}$ by Ca$^{2+}$, Ca$_{2}$CrReO$_{6}$ becomes a FM insulator with $T_{\rm C}$ of 360 K\cite{Sr2CrReO6}, while Ca$_{2}$CrOsO$_{6}$ remains to be a FM insulator with reduced $T_{\rm C}$ of 490 K\cite{Ca2CrOsO6}. In addition, the novel magnetism observed in Sr$_{2}$YIrO$_{6}$ and Ba$_{2}$YIrO$_{6}$ has raised discussions about the Ir$^{5+}$ $j_{\rm eff}=0$ state\cite{Sr2YIrO6,Ba2YIrO6_2016,Ba2YIrO6_2017}.

\begin{figure}[t]
  \centering
\includegraphics[width=8.5cm]{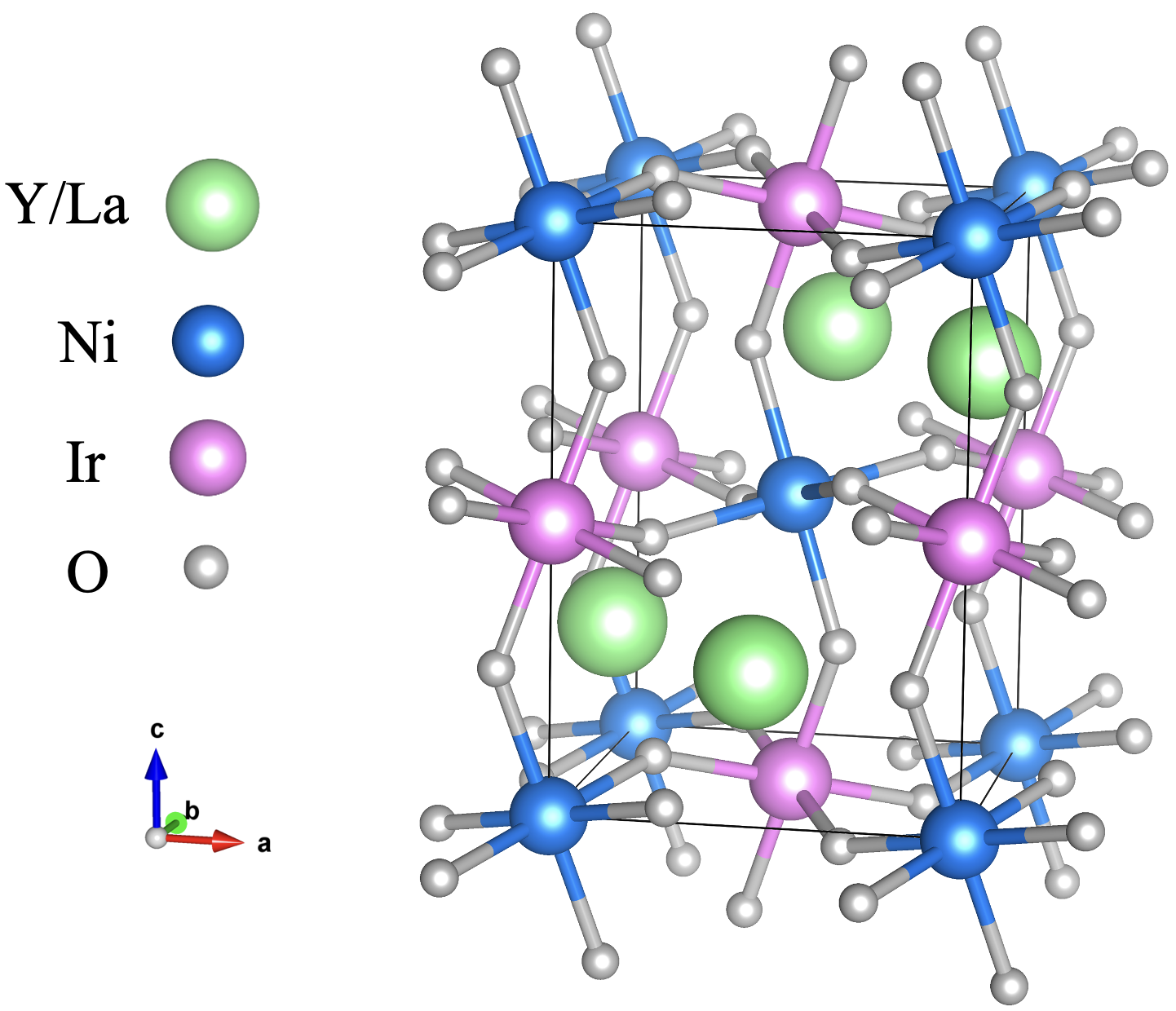}
  \caption{Crystal structure of the double perovskites $A_{2}$NiIrO$_{6}$ ($A$=Y, La): the Ni and Ir ions form their respective distorted fcc sublattices.}
  \label{fig:1}
\end{figure}

Very recently, double perovskite Y$_{2}$NiIrO$_{6}$ has been synthesized, adopting the Ni-Ir rock-salt ordered structure with the monoclinic space group $P2_{1}/n$\cite{Y_2023}, as shown in Fig. \ref{fig:1}. X-ray absorption spectroscopy confirms the Ni$^{2+}$ and Ir$^{4+}$ valence states. Moreover, the Ni$^{2+}$ and Ir$^{4+}$ ions are found to be AFM coupled below the $T_{\rm N}$ of 192 K. As an analog to Y$_{2}$NiIrO$_{6}$, La$_{2}$NiIrO$_{6}$ also contains the Ni$^{2+}$ and Ir$^{4+}$ ions and crystallizes in the $P2_{1}/n$ structure\cite{La_acta2021,La_PRM2021,La_PRM2022}. On the contrary, the $T_{\rm N}$ of La$_{2}$NiIrO$_{6}$ is significantly reduced to 74-80 K\cite{La_acta2021,La_PRM2021,La_PRM2022}. Owing to the much smaller ionic size of Y$^{3+}$ than La$^{3+}$, Y$_{2}$NiIrO$_{6}$ is more distorted than La$_{2}$NiIrO$_{6}$, with the more bent Ni-O-Ir bonds in the former than the latter, being 139.9$\degree$-142.6$\degree$ for the bond angles in Y$_{2}$NiIrO$_{6}$ and 151.6$\degree$-153.8$\degree$ in La$_{2}$NiIrO$_{6}$. It is often that more bent bonds could yield a weaker magnetic coupling\cite{Khomskii_2001}, but surprisingly here the more distorted Y$_{2}$NiIrO$_{6}$ has a much higher $T_{\rm N}$ than La$_{2}$NiIrO$_{6}$\cite{La_acta2021,La_PRM2021,La_PRM2022}. The exchange mechanism responsible for the magnetic behaviors remains a matter of debate\cite{Y_2023,La_acta2021,La_PRM2021,La_PRM2022, La_PRB2022}.

\begin{figure}[t]
  \centering
\includegraphics[width=8.5cm]{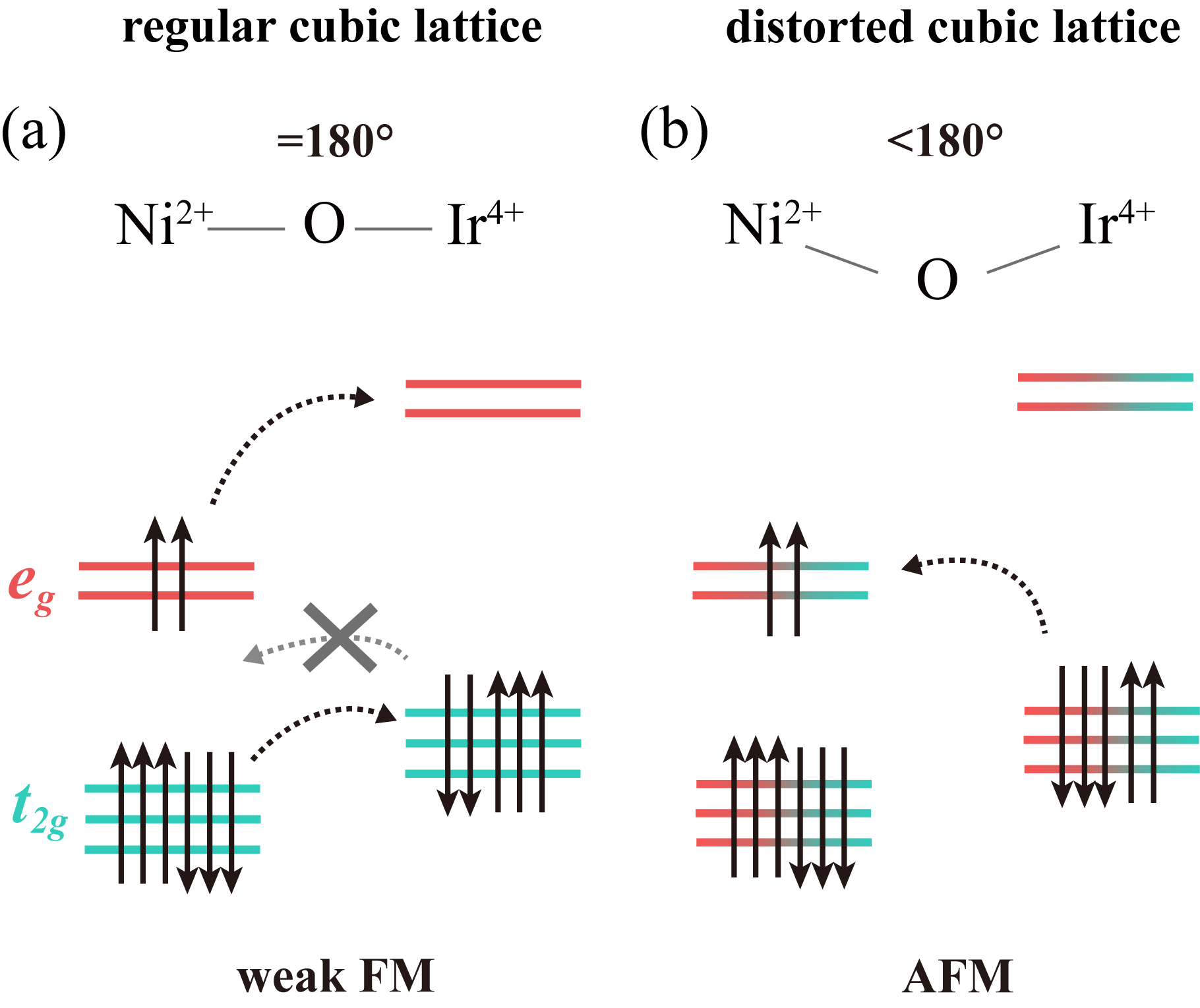}
  \caption{Schematic crystal field level diagrams of the Ni$^{2+}$ and Ir$^{4+}$ ions, and the possible superexchange channels. (a) In the regular cubic lattice, the normal excitation from the Ni$^{2+}$-Ir$^{4+}$ to the intermediate Ni$^{+}$-Ir$^{5+}$ is forbidden, but the abnormal (unusual) excitation to the Ni$^{3+}$-Ir$^{3+}$ with a large excitation gap would yield a weak FM coupling. (b) The normal excitation to Ni$^{+}$-Ir$^{5+}$, associated with the mixed $t_{2g}$-$e_g$ states in the distorted cubic lattice, gives the AFM coupling, see more in the main text.}
  \label{fig:2}
\end{figure}

In this paper, we study the electronic structure and varying magnetism in Y$_{2}$NiIrO$_{6}$ and La$_{2}$NiIrO$_{6}$. Starting from the Ni$^{2+}$-Ir$^{4+}$ state as confirmed below, and if the structural details would not be considered, an assumed regular cubic lattice would only give a weak FM Ni-Ir coupling according to the crystal field level diagrams and the superexchange pictures, see Fig.~\ref{fig:2}(a). Obviously, this contradicts with the experimental AFM order, and therefore, the lattice distortion should play a vital role in determining the AFM order. As depicted in Fig.~\ref{fig:2}(b), the  mixed $t_{2g}$-$e_g$ states due to the lattice distortion allow the electron hopping from the Ir $t_{2g}$ to Ni $e_g$ and thus facilitate the AFM coupling. As confirmed below by our first-principles calculations and Monte Carlo simulations, it is the lattice distortion induced $t_{2g}$-$e_g$ mixing that gives a stronger Ni-Ir AFM coupling and thus higher $T_{\rm N}$ in the more distorted Y$_{2}$NiIrO$_{6}$ than in La$_{2}$NiIrO$_{6}$. Moreover, our calculations indeed show that Y$_{2}$NiIrO$_{6}$ in a fictitious cubic structure would instead have the much weaker FM coupling. Furthermore, we find that the Ir$^{4+}$ ion is not in the $j_{\rm eff}=1/2$ state due to the $t_{2g}$ level splitting, Hund exchange, and the band formation, but that it is in the $S=1/2$ state and carries a finite orbital moment via SOC.

\section{Computational Details}
Density functional theory calculations are carried out using the full-potential augmented plane wave plus local orbital code (WIEN2k)\cite{WIEN2k}. The structural relaxation is carried out using the local spin density approximation (LSDA). The optimized lattice parameters of $a$=5.207 \AA, $b$=5.631 \AA, $c$=7.499 \AA, and $\beta$=89.94$\degree$ for Y$_{2}$NiIrO$_{6}$ are almost the same (within 1$\%$) as the experimental ones of $a$=5.260 \AA, $b$=5.689 \AA, $c$=7.576 \AA, and $\beta$=90.16$\degree$\cite{Y_2023}. For La$_{2}$NiIrO$_{6}$, the optimized lattice parameters of $a$=5.472 \AA, $b$=5.535 \AA, $c$=7.754 \AA, and $\beta$=89.91$\degree$ are also close to (within 1.7$\%$) the experimental ones of $a$=5.566 \AA, $b$=5.630 \AA, $c$=7.888 \AA, and $\beta$=90.09$\degree$\cite{La_PRM2021}. Consistent with the experimental structures, our optimized structures confirm a stronger lattice distortion in Y$_{2}$NiIrO$_{6}$ than in La$_{2}$NiIrO$_{6}$, as evidenced by the more bent Ni-O-Ir bond angles (ranging from 140.2$\degree$ to 142.9$\degree$) in the former than those (150.3$\degree$-152.0$\degree$) in the latter. The muffin-tin sphere radii are chosen to be 2.6, 2.1, 2.1, and 1.5 bohrs for Y/La, Ni, Ir, and O atoms, respectively. The cutoff energy of 16 Ry is used for plane wave expansion, and 400 k-points are sampled for integration over the first Brillouin zone. To account for the electron correlation effect of Ni 3$d$ and Ir 5$d$ electrons, we employ a hybrid functional with a quarter Hartree-Fock exchange mixed into LSDA\cite{Becke_1993,Becke_1996, Fock25, hf_1,hf_2}. We also test the LSDA plus Hubbard $U$ (LSDA+$U$) method\cite{Anisimov_1993} with the common $U=$ 5 eV (3 eV) and the Hund exchange $J_{\rm H}=$ 1 eV (0.4 eV) for Ni 3$d$ (Ir 5$d$) electrons\cite{JH_1,JH_2,JH_3}. Both sets of the calculations give similar results as seen below. Spin orbit coupling (SOC) is included by the second variational method with scalar relativistic wave functions, and the magnetization direction is set along the $c$-axis. To estimate the magnetic transition temperature, Monte Carlo simulations using the Metropolis method\cite{Nicholas_1949} have been performed on a 12$\times$12$\times$12 spin cell. At each temperature, 4.8$\times$10$^{7}$ spin flips are performed to reach an equilibrium. The magnetization is sampled after 1.3$\times$10$^{4}$ spin flips, and 2$\times$10$^{4}$ magnetizations are used to take the average.

\renewcommand\arraystretch{1.2}
\begin{table*}[t]
 \caption{Relative total energies $\Delta E$ (meV/fu) and local spin moments ($\mu_{\rm B}$) in different magnetic states by hybrid functional calculations. The corresponding LSDA+$U$ results are summarized in the brackets. The derived exchange parameters (meV) are listed in the last column.}
  \label{tb1}
\begin{ruledtabular}
  \setlength{\tabcolsep}{0.1mm}{
  \begin{tabular}{ccrrrl}
 & States & $\Delta E$ & Ni$^{2+}$  &  Ir$^{4+}$ & $J$ \\ \hline
Y$_{2}$NiIrO$_{6}$  &AFM  & 0.0 (0.0) & 1.66 (1.59)   & $-$0.52 ($-$0.48) & \multirow{2}*{$J_{\rm Ni-Ir}=$ 18.27 (16.90)}    \\
   &FM    & 109.6 (101.4)  & 1.72 (1.65)  & 0.63 (0.61) \\
   
Y$_{2}$NiGeO$_{6}$  &layered-AFM  & 0.0 (0.0) & $\pm$1.71 ($\pm$1.66)   & / & \multirow{2}*{$J^{\prime}_{\rm Ni-Ni}=$ 0.35 (0.35)}     \\
  &FM    & 2.8 (2.8)  & 1.72 (1.66)   & /  \\
  
Y$_{2}$ZnIrO$_{6}$  &layered-AFM  & 0.0 (0.0) & /   &$\pm$0.58 ($\pm$0.53) &\multirow{2}*{$J^{\prime}_{\rm Ir-Ir}=$ 1.40 (1.10)} \\
&FM    & 2.8 (2.2)  & /   & 0.60 (0.56)  \\                 \hline

La$_{2}$NiIrO$_{6}$  &AFM  & 0.0 (0.0) & 1.69 (1.62)   & $-$0.51 ($-$0.47) & \multirow{2}*{$J_{\rm Ni-Ir}=$ 11.72 (12.85)} \\
   &FM    & 70.3 (77.1)  & 1.74 (1.67)  & 0.66 (0.64) \\
   
La$_{2}$NiGeO$_{6}$  &layered-AFM  & 0.0 (0.0) & $\pm$1.70 ($\pm$1.66)  & / & \multirow{2}*{$J^{\prime}_{\rm Ni-Ni}=$ 1.69 (1.03)}     \\
  &FM    &13.5 (8.2)   & 1.71 (1.66)   & /  \\
  
La$_{2}$ZnIrO$_{6}$  &layered-AFM  & 0.0 (0.0) & /   &$\pm$0.58 ($\pm$0.55) &\multirow{2}*{$J^{\prime}_{\rm Ir-Ir}=$ $-$0.65 ($-$1.95)}  \\
&FM    & $-$1.3 ($-$3.9)  & /   & 0.59 (0.56)  \\                 \hline

\multirow{2}*{\parbox[c]{2.7cm}{Y$_{2}$NiIrO$_{6}$ (artificial cubic structure)}} &AFM  & 0.0 (0.0)  & 1.72 (1.69)  & $-$0.54 ($-$0.52) &\multirow{2}*{$J_{\rm Ni-Ir}=$ $-$4.70 ($-$4.03)} \\
 &FM    & $-$28.2 ($-$24.2)  & 1.75 (1.72)   & 0.73 (0.71)  \\
       
 \end{tabular}}
 \end{ruledtabular}
\end{table*}

\begin{figure}[t]
  \centering
\includegraphics[width=8.5cm,height=7.7cm]{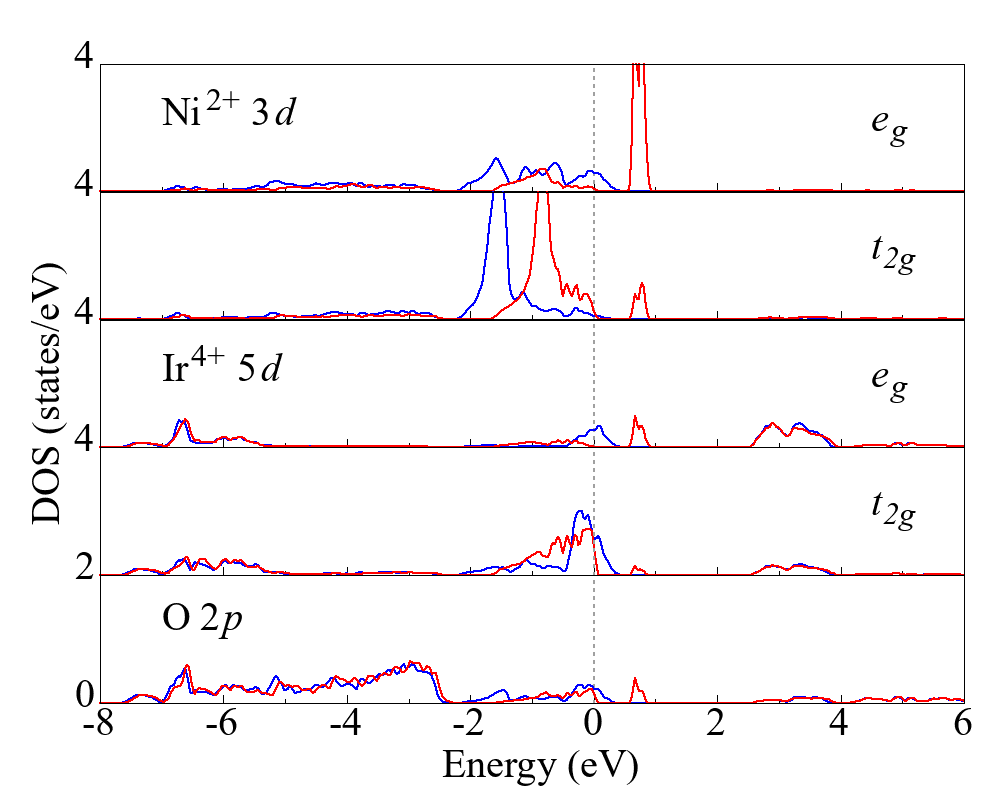}
  \caption{The Ni$^{2+}$ $3d$, Ir$^{4+}$ $5d$, and O $2p$ DOS for Y$_{2}$NiIrO$_{6}$ by LSDA. The blue (red) curves stand for the up (down) spins. The Fermi level is set at zero energy.}
  \label{fig:3}
\end{figure}

\section{Results and discussion}
We first carry out LSDA calculations to see spin polarization, valence states, and crystal field in Y$_{2}$NiIrO$_{6}$. The LSDA calculations naturally converge to the Ni-Ir AFM state. To obtain the FM state, we perform fixed-spin-moment calculations, and find that the FM-fixed state with 3 $\mu_{\rm B}$/fu (Ni$^{2+}$ $S$=1 plus Ir$^{4+}$ $S$=1/2) is less stable than the AFM one by 74.8 meV/fu. We plot the orbitally resolved density of states (DOS) for the AFM state in Fig. \ref{fig:3}. For the Ni 3$d$ states, the octahedral crystal field gives rise to the $e_{g}$-$t_{2g}$ splitting of about 1.5 eV, and only the down-spin $e_{g}$ states are unoccupied, giving the formal Ni$^{2+}$ charge state with the high-spin $t_{2g}$$^{6}$$e_{g}$$^{2}$ ($S=1$) configuration. The Ni 3$d$-O 2$p$ hybridization yields some Ni 3$d$ states below $-$2 eV, and this, together with the Ni-Ir AFM couplings, reduces the local spin moment of the Ni$^{2+}$ ion to 1.18 $\mu_{\rm B}$. As for Ir 5$d$ electrons, the crystal field and strong Ir 5$d$-O 2$p$ hybridization result in a large $e_{g}$-$t_{2g}$ splitting of about 3.5 eV, leaving the unoccupied $e_{g}$ states lying at 2.5-4 eV above the Fermi level. The fully occupied down-spin $t_{2g}$ states and partially occupied up-spin $t_{2g}$ ones imply the formal Ir$^{4+}$ valence state with the low-spin $t_{2g}$$^{5}$ ($S=1/2$) configuration. However, the Ir$^{4+}$ ion has the largely reduced spin moment of $-$0.19 $\mu_{\rm B}$ due to the strong band hybridization. It is important to note that, owing to the lattice distortion from the ideal cubic structure, there exist apparent mixtures of the $e_{g}$ and $t_{2g}$ orbitals for both the Ni 3$d$ and Ir 5$d$ states, see Fig. \ref{fig:3}.

\begin{figure}[t]
  \centering
\includegraphics[width=8.5cm]{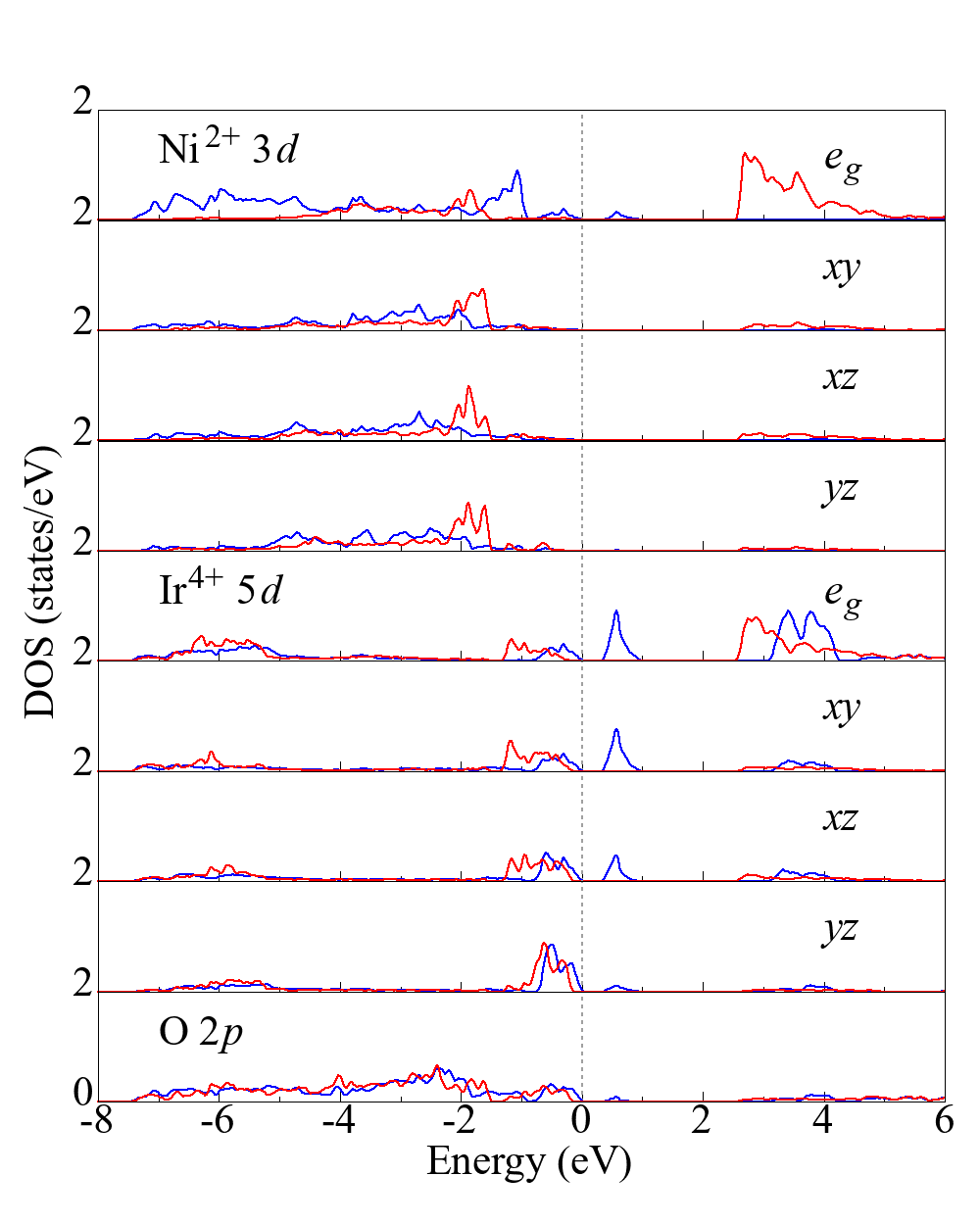}
  \caption{DOS of the Ni$^{2+}$ $3d$, Ir$^{4+}$ $5d$, and O $2p$ states for Y$_{2}$NiIrO$_{6}$ by hybrid functional calculations. The blue (red) curves stand for the up (down) spins. The Fermi level is set at zero energy.}
  \label{fig:4}
\end{figure}

\begin{figure}[t]
  \centering
\includegraphics[width=8.5cm]{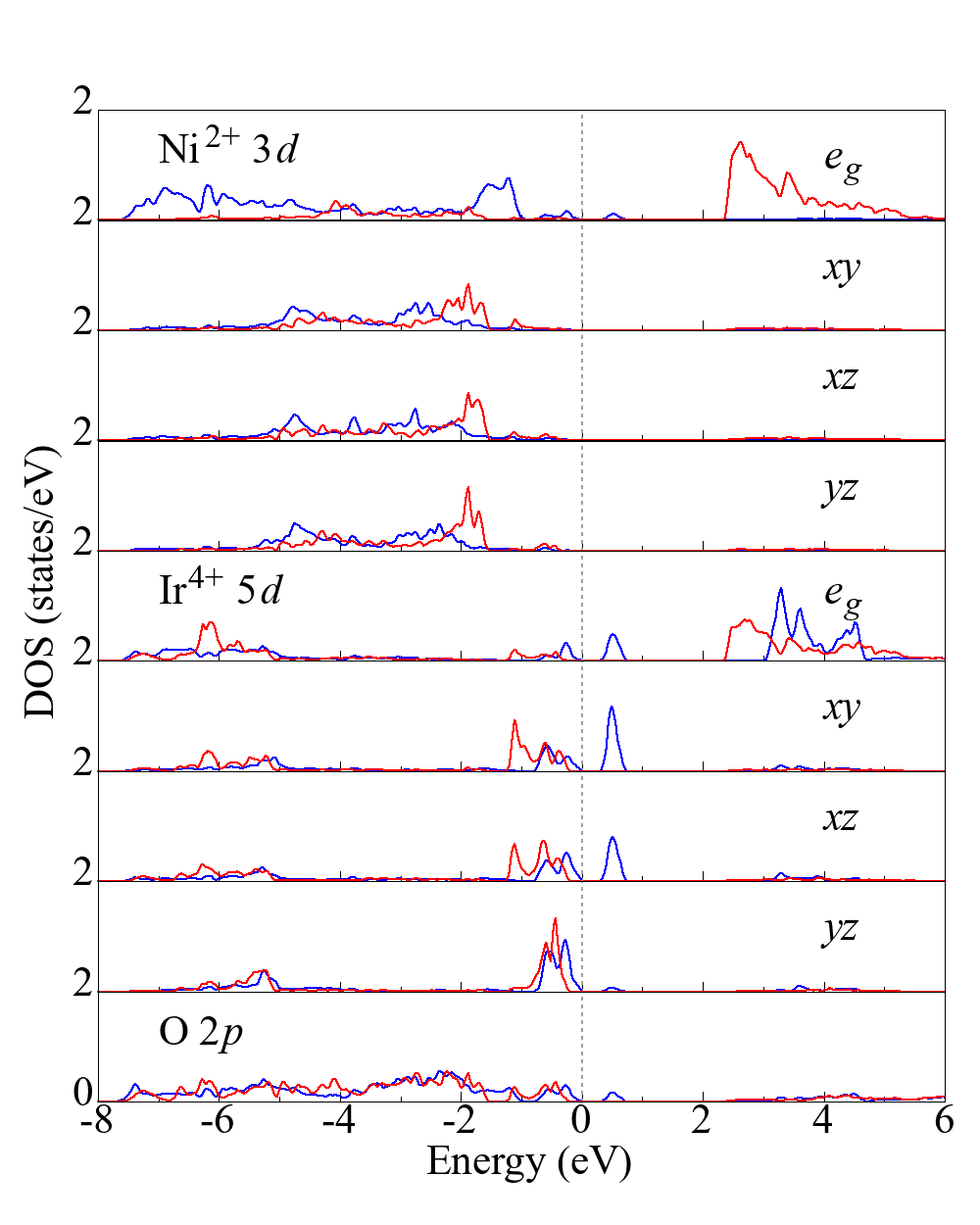}
  \caption{DOS of the Ni$^{2+}$ $3d$, Ir$^{4+}$ $5d$, and O $2p$ states for La$_{2}$NiIrO$_{6}$ by hybrid functional calculations. The blue (red) curves stand for the up (down) spins. The Fermi level is set at zero energy.}
  \label{fig:5}
\end{figure}

The above LSDA calculations give a metallic solution for Y$_2$NiIrO$_6$, and this is probably due to an overestimation of electron delocalization within this framework. To compensate for this effect, we carry out hybrid functional calculations. The Ni$^{2+}$ and Ir$^{4+}$ states are now localized in an insulating solution with the inclusion of electron correlation effect. We present the DOS results of the AFM ground state in Fig. \ref{fig:4}. The Ni 3$d$ states have a large gap of about 3.5 eV, and the unoccupied down-spin Ni $e_{g}$ states lie at 2.5-5 eV above the Fermi level due to the strong correlation effect. The Ni$^{2+}$ $t_{2g}$$^{6}$$e_{g}$$^{2}$ ($S=1$) configuration is further stabilized by a comparison with that in LSDA, and its local spin moment increases up to 1.66 $\mu_{\rm B}$. In contrast, the relatively weak correlation effect of Ir 5$d$ electrons opens a small insulating gap of about 0.3 eV for Ir $t_{2g}$ bands, where the single $t_{2g}$ hole state is mainly composed of the up-spin $xy$ orbital. As a result, the Ir$^{4+}$ ion is in the $t_{2g}$$^{5}$ ($S=1/2$) state and has a local spin moment of $-$0.52 $\mu_{\rm B}$. In addition, we note that the non-negligible $t_{2g}$-$e_{g}$ orbital mixing persists due to the lattice distortion, see Figs. \ref{fig:3} and \ref{fig:4}.

For Y$_2$NiIrO$_6$ in the charge-spin-orbital states of Ni$^{2+}$ $t_{2g}$$^{6}$$e_{g}$$^{2}$ ($S=1$) and Ir$^{4+}$ $t_{2g}$$^{5}$ ($S=1/2$), both the AFM and FM states are stabilized in the hybrid functional calculations. We again obtain the AFM ground state, and it is more stable than the FM state by 109.6 meV/fu, see Table \ref{tb1}. Then the first nearest-neighboring (1NN) Ni$^{2+}$-Ir$^{4+}$ AFM exchange parameter is estimated to be $J_{\rm Ni-Ir}=(E_{\rm FM}-E_{\rm AFM})/12S_{\rm Ni^{2+}}S_{\rm Ir^{4+}}=18.27$ meV, counting the magnetic exchange energy of each Ni-Ir pair by $J_{\rm Ni-Ir}S_{\rm Ni^{2+}}S_{\rm Ir^{4+}}$ and the six coordination of the Ni-Ir ions. In this estimate, we assume all the 1NN Ni-Ir couplings to be equal by neglecting their possible variation associated with the delicate structural details. As discussed in the Introduction and depicted in Fig. \ref{fig:2}(b), the AFM Ni$^{2+}$-Ir$^{4+}$ coupling is given by the Ir$^{4+}$ $t_{2g}$-O $2p$-Ni$^{2+}$ $e_{g}$ superexchange with assistance of the $t_{2g}$-$e_{g}$ orbital mixing due to the lattice distortion. This orbital mixing allows, otherwise forbids, the virtual electron hopping from the Ir$^{4+}$ $t_{2g}$ to the Ni$^{2+}$ $e_{g}$ in the charge fluctuation process from the Ir$^{4+}$-Ni$^{2+}$ ground state to the intermediate Ir$^{5+}$-Ni$^{1+}$ excited state (with a much lower energy cost than the excitation to the Ir$^{3+}$-Ni$^{3+}$ state which is less common or even rare).

Now we consider the possible far-distance magnetic interactions which may be important in perovskite oxides containing the delocalized 5$d$ electrons\cite{ou_2014,ke_2022,hou_2015}. To estimate the 2NN Ni$^{2+}$-Ni$^{2+}$ and Ir$^{4+}$-Ir$^{4+}$ exchange parameters, we use two artificial systems Y$_{2}$NiGeO$_{6}$ and Y$_{2}$ZnIrO$_{6}$ both in the Y$_2$NiIrO$_6$ structure, assuming a substitution of nonmagnetic Ge$^{4+}$ for Ir$^{4+}$ ions (both in the same valence state), or nonmagnetic Zn$^{2+}$ for Ni$^{2+}$ ions. This approach avoids using of bigger supercells for complicated magnetic structures and enables us to estimate the Ni-Ni and Ir-Ir magnetic couplings ($J'_{\rm Ni-Ni}$ and $J'_{\rm Ir-Ir}$) separately. We set the layered-AFM state with the FM coupling in the $ab$ plane and the AFM one along the $c$ axis. This layered-AFM state and FM state differs only by the 2NN Ni$^{2+}$-Ni$^{2+}$ (Ir$^{4+}$-Ir$^{4+}$) exchange for Y$_{2}$NiGeO$_{6}$ (Y$_{2}$ZnIrO$_{6}$). As shown in Table \ref{tb1}, the FM state of Y$_{2}$NiGeO$_{6}$ lies higher than the layered-AFM one by 2.8 meV/fu, then the AFM $J'_{\rm Ni-Ni}$ parameter is estimated to be $J'_{\rm Ni-Ni}=(E_{\rm FM}-E_{\rm layered-AFM})/8S_{\rm Ni^{2+}}S_{\rm Ni^{2+}}=0.35$ meV. Here again, we assume all the 2NN Ni-Ni couplings to be equal by neglecting the delicate structural details. Similarly, we obtain the AFM $J'_{\rm Ir-Ir}$ = 1.40 meV, see Table \ref{tb1}. Now we see that the 2NN $J'_{\rm Ni-Ni}$ and $J'_{\rm Ir-Ir}$ are at least one order of magnitude smaller than the 1NN $J_{\rm Ni-Ir}$, and therefore, the AFM behavior of Y$_{2}$NiIrO$_{6}$ would be dominated by the 1NN Ni-Ir couplings.

So far, we have found that Y$_{2}$NiIrO$_{6}$ is an AFM Mott insulator with the dominant Ni$^{2+}$-Ir$^{4+}$ AFM coupling, which is in agreement with the experimental AFM order\cite{Y_2023}. We have also performed LSDA+$U$ calculations, and they give quite similar results as hybrid functional calculations: the AFM state for Y$_{2}$NiIrO$_{6}$ by LSDA+$U$ is more stable than the FM one by 101.4 meV/fu, and the derived 1NN Ni$^{2+}$-Ir$^{4+}$ AFM parameter $J_{\rm Ni-Ir}$ = 16.90 meV turns out to be one order of magnitude stronger than the 2NN Ni$^{2+}$-Ni$^{2+}$ and Ir$^{4+}$-Ir$^{4+}$ couplings, see Table \ref{tb1}. Thus, both the hybrid functional calculations and LSDA+$U$ ones give the consistent results reproducing the AFM Mott insulating behavior of Y$_{2}$NiIrO$_{6}$.

Now we turn to La$_{2}$NiIrO$_{6}$ and find that LSDA calculations give a very similar AFM metallic solution (not shown here) to that for Y$_{2}$NiIrO$_{6}$ (see Fig. \ref{fig:3}). However, the Ni-Ir AFM coupling seems much weaker in La$_{2}$NiIrO$_{6}$ than Y$_{2}$NiIrO$_{6}$, as implied by the LSDA results that the energy difference between the FM and AFM states is reduced to 15.4 meV/fu for La$_{2}$NiIrO$_{6}$, being much smaller than that of 74.8 meV/fu for Y$_{2}$NiIrO$_{6}$ (see above). To restore the Mott insulating behavior of La$_{2}$NiIrO$_{6}$ and the magnetic superexchange, we perform both the hybrid functional and LSDA+$U$ calculations both to include electron correlation effects, and they turn out to give quite similar results. As such, we focus on the hybrid functional calculations as shown in Fig. \ref{fig:5}. La$_{2}$NiIrO$_{6}$ is an AFM Mott insulator with an insulating gap of about 0.3 eV in good agreement with the previous experiments~\cite{La_PRB2022}, and all the electronic structure closely resembles that of Y$_{2}$NiIrO$_{6}$, see Figs. \ref{fig:4} and \ref{fig:5} for a comparison. Then, La$_{2}$NiIrO$_{6}$ is in the Ni$^{2+}$ $t_{2g}$$^{6}$$e_{g}$$^{2}$ ($S=1$) and Ir$^{4+}$ $t_{2g}$$^{5}$ ($S=1/2$) state, too. The band hybridization and the AFM Ni-Ir coupling both lead to the reduced local spin moments of 1.69 $\mu_{\rm B}$/Ni$^{2+}$ and $-$0.51 $\mu_{\rm B}$/Ir$^{4+}$, see Table \ref{tb1}. We also estimate the exchange parameters for the AFM La$_{2}$NiIrO$_{6}$ and find that 1NN AFM $J_{\rm Ni-Ir}$ = 11.72 meV is again about one order of magnitude stronger than the 2NN $J'_{\rm Ni-Ni}$ = 1.69 meV and $J'_{\rm Ir-Ir}$ = --0.65 meV. As the 1NN AFM $J_{\rm Ni-Ir}$ is dominant, we would not go into the details to discuss the 2NN Ni-Ni and Ir-Ir couplings. Here, more attention is paid to the 1NN Ni-Ir coupling, which is AFM type due to the superexchange associated with the $t_{2g}$-$e_g$ orbital mixing due to the lattice distortion, see Fig. \ref{fig:2}(b).

\begin{figure}[t]
  \centering
\includegraphics[width=7.5cm]{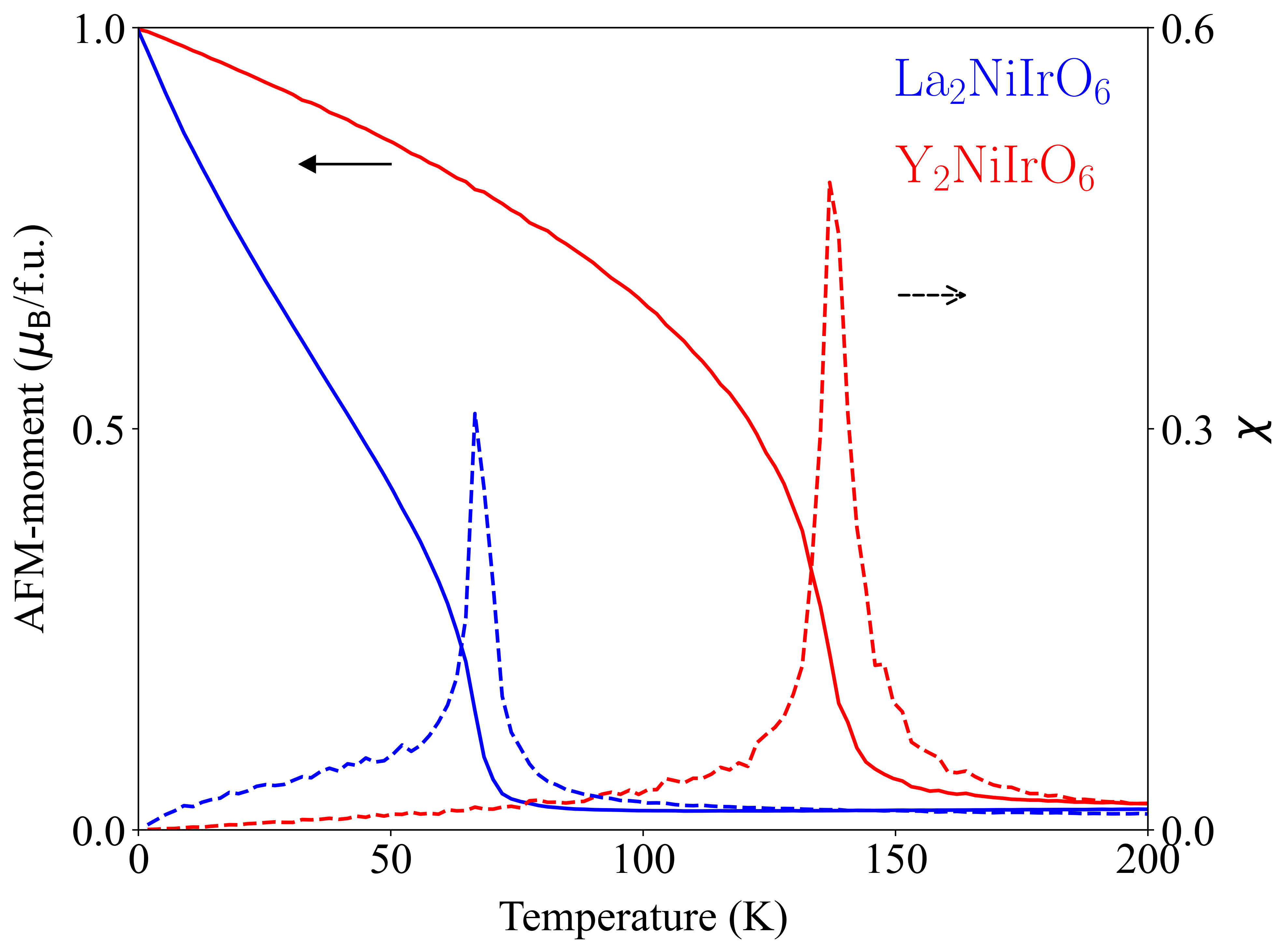}
  \caption{Monte Carlo simulations of the magnetization and magnetic susceptibility in Y$_{2}$NiIrO$_{6}$ and La$_{2}$NiIrO$_{6}$.}
  \label{fig:6}
\end{figure}

Note that in an ideal cubic lattice, the $t_{2g}$ and $e_g$ states are orthorgonal, and they cannot be mixed at all. But $t_{2g}$-$e_g$ orbital mixing is allowed by a lattice distortion. As Y$_{2}$NiIrO$_{6}$ has a larger lattice distortion and thus stronger Ni-O-Ir bond bending than La$_{2}$NiIrO$_{6}$, the $t_{2g}$-$e_g$ orbital mixing is stronger in the former than in the latter, and this is clearly seen, e.g., in the Ir $5d$ DOS results shown in Figs. \ref{fig:4} and \ref{fig:5} (more specifically, seeing the single `$t_{2g}$' hole state mixed with the `$e_g$' at 0.5 eV above the Fermi level). Therefore, it is natural that the 1NN AFM Ni-Ir coupling is stronger in Y$_{2}$NiIrO$_{6}$ than in La$_{2}$NiIrO$_{6}$, which is indeed confirmed by our hybrid functional (and LSDA+$U$) calculations giving $J_{\rm Ni-Ir}$ = 18.27 meV (16.90 meV) for Y$_{2}$NiIrO$_{6}$ and 11.72 meV (12.85 meV) for La$_{2}$NiIrO$_{6}$ as seen in Table \ref{tb1}. Then, using the dominant 1NN $J_{\rm Ni-Ir}$ and the one order of magnitude weaker 2NN $J'_{\rm Ni-Ni}$ and $J'_{\rm Ir-Ir}$ given by the hybrid functional calculations as summarized in Table \ref{tb1}, we estimate $T_{\rm N}$ through Monte Carlo simulations based on the spin Hamiltonian
\begin{align*}
H&= \sum_{\langle i, j\rangle}\frac{J_{\rm Ni-Ir}}{2} \bm{S}_{i}^{\rm Ni}\cdot \bm{S}_{j}^{\rm Ir}\\
&+\sum_{\langle\langle i, j\rangle\rangle} (\frac{J^{\prime}_{\rm Ni-Ni}}{2}  \bm{S}_{i}^{\rm Ni}\cdot \bm{S}_{j}^{\rm Ni}+\frac{J^{\prime}_{\rm Ir-Ir}}{2} \bm{S}_{i}^{\rm Ir}\cdot \bm{S}_{j}^{\rm Ir}).
\end{align*}
The $T_{\rm N}$ is estimated to be 138 K for Y$_{2}$NiIrO$_{6}$ and 67 K for La$_{2}$NiIrO$_{6}$, see Fig. \ref{fig:6}. Although our computational $T_{\rm N}$ values have a quantitative difference from the experimental results, our above results and analyses are in line with the experimental findings that the $T_{\rm N}$ is significantly increased from 74-80 K for La$_{2}$NiIrO$_{6}$ to 192 K for Y$_{2}$NiIrO$_{6}$\cite{Y_2023,La_acta2021,La_PRM2021,La_PRM2022}.

Our above results and analyses show that the $t_{2g}$-$e_g$ orbital mixing due to the lattice distortion facilitates the AFM Ni-Ir coupling, which is depicted in Fig. \ref{fig:2}(b). To further verify this picture, we assume an ideal cubic structure for Y$_{2}$NiIrO$_{6}$ as a counterpart to see its possible magnetic structure. In the cubic lattice, there is no $t_{2g}$-$e_g$ orbital mixing, and thus the virtual hopping plotted in Fig. \ref{fig:2}(b) is strictly forbidden. Moreover, owing to the full filling of the Ni $t_{2g}$ orbitals, the Ir $t_{2g}$ electrons cannot hop at all to the Ni $t_{2g}$ in the normal charge fluctuation process from the Ir$^{4+}$-Ni$^{2+}$ ground state to the intermediate Ir$^{5+}$-Ni$^{1+}$ excited state. In sharp contrast, the unusual (abnormal) charge fluctuation into Ir$^{3+}$-Ni$^{3+}$ state has to be invoked, as    plotted in Fig. \ref{fig:2}(a). Then both the up-spin Ni $e_g$ and down-spin $t_{2g}$ electrons could virtually hop to the Ir site to form a FM superexchange. However, as the empty Ir $e_g$ state is quite high (about 3 eV above the Fermi level, see Fig. \ref{fig:3}), the energy cost to reach it is too large, and the excited Ir$^{3+}$-Ni$^{3+}$ intermediate state is less common or even rare. Therefore, the FM superexchange would be weak in the fictitious cubic lattice, compared with the AFM one in the real distorted lattice as demonstrated above. Indeed, our hybrid functional calculations and LSDA+$U$ ones both consistently show that Y$_{2}$NiIrO$_{6}$ in the fictitious cubic structure would instead be in the FM ground state, and that here the FM Ni-Ir exchange strength is only a quarter of the AFM Ni-Ir coupling in the real lattice, see Table \ref{tb1}. Thus, our picture seems to be well established, as seen in Fig. \ref{fig:2}. It is the lattice distortion which plays the vital role in determining the AFM structure and the significant $T_{\rm N}$ increase from the less distorted La$_{2}$NiIrO$_{6}$ to the more distorted Y$_{2}$NiIrO$_{6}$.

\begin{figure}[t]
  \centering
\includegraphics[width=8.5cm]{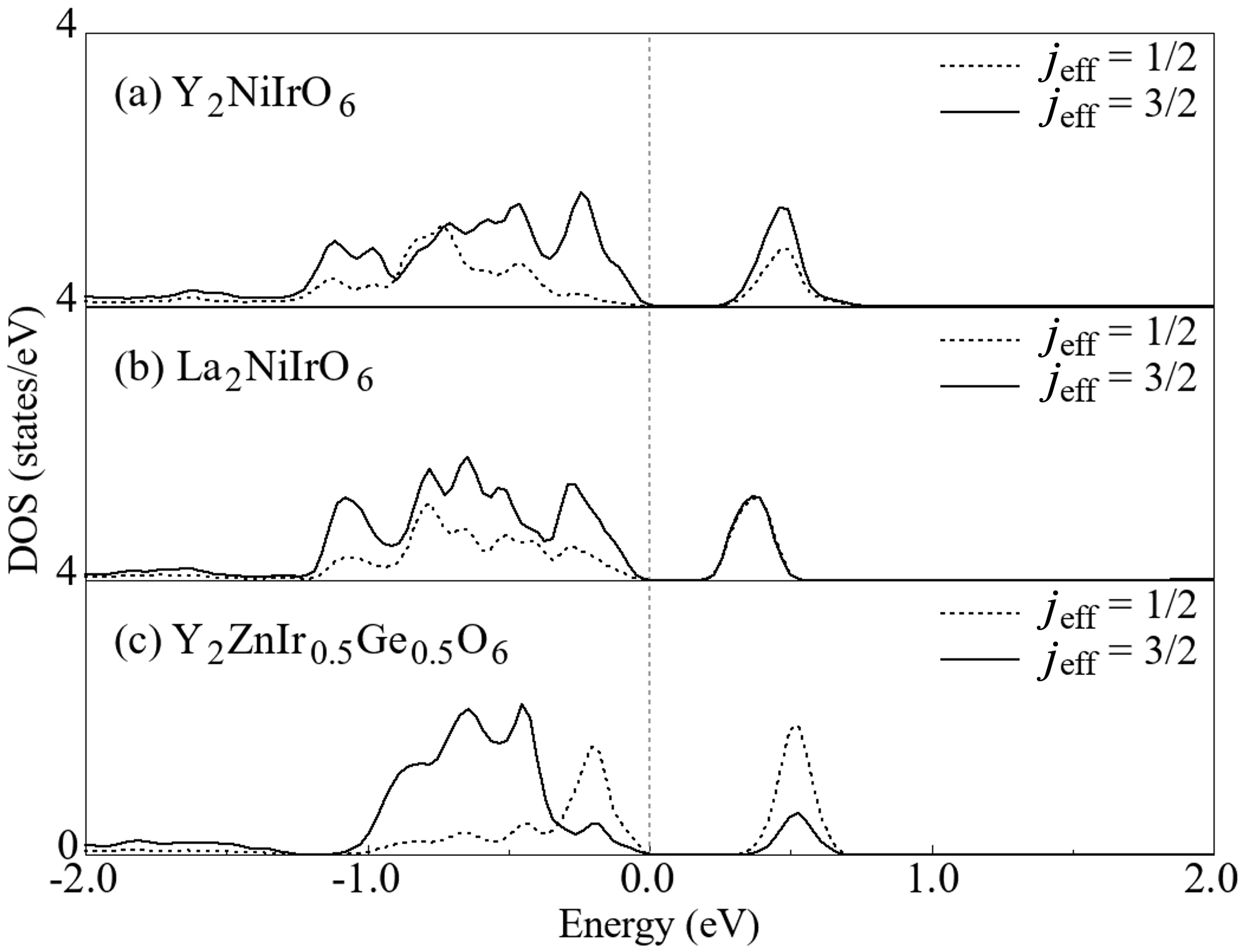}
  \caption{The hybrid functional+SOC calculated Ir$^{4+}$ $t_{2g}$$^{5}$ DOS projected onto the $j_{\rm eff}=1/2$ and $j_{\rm eff}=3/2$ basis for (a) Y$_{2}$NiIrO$_{6}$, (b) La$_{2}$NiIrO$_{6}$, and (c) Y$_{2}$ZnIr$_{0.5}$Ge$_{0.5}$O$_{6}$. The significant mixing of $j_{\rm eff}=1/2$ and $j_{\rm eff}=3/2$ in Y$_{2}$NiIrO$_{6}$ and La$_{2}$NiIrO$_{6}$ is largely suppressed in Y$_{2}$ZnIr$_{0.5}$Ge$_{0.5}$O$_{6}$ which seems to restore the $j_{\rm eff}=1/2$ state.}
  \label{fig:7}
\end{figure}

Finally, we check how SOC acts in Y$_{2}$NiIrO$_{6}$ and La$_{2}$NiIrO$_{6}$ as SOC is often of concern for heavy elements like Ir atoms. The large $e_{g}$-$t_{2g}$ splitting of the Ir $5d$ orbitals leaves the $e_{g}$ doublet out of consideration, and the $t_{2g}$ levels would split by SOC into the higher doublet $j_{\rm eff}=1/2$ and the lower quartet $j_{\rm eff}=3/2$. Then the Ir$^{4+}$ $t_{2g}$$^{5}$ configuration may be in the well-known $j_{\rm eff}=1/2$ state. This was initially proposed for Sr$_{2}$IrO$_{4}$\cite{Kim_2008PRL} and further extended for many other $5d$ and $4d$ TM compounds\cite{takagi_2019,clark_2021}. If the $j_{\rm eff}=1/2$ picture is adopted for Ir$^{4+}$ ions in Y$_{2}$NiIrO$_{6}$ and La$_{2}$NiIrO$_{6}$, in combination with the Ni$^{2+}$ $S=1$ state, the effective magnetic moment $\mu_{\rm eff}$ is expected to be the same as that of the Ir$^{4+}$ $S$ =1/2 and Ni$^{2+}$ $S$ = 1 state, i.e., $\sqrt{g_{S}^{2}S_{\rm Ir}(S_{\rm Ir}+1)+g_{S}^{2}S_{\rm Ni}(S_{\rm Ni}+1)}=\sqrt{4\times\frac{1}{2}\times\frac{3}{2}+4\times1\times2}\approx 3.32$ $\mu_{\rm B}$/fu which is an atomic upper limit. However, the experimental $\mu_{\rm eff}$ is 3.85 $\mu_{\rm B}$/fu for Y$_{2}$NiIrO$_{6}$\cite{Y_2023} and 3.38-3.84 $\mu_{\rm B}$/fu for La$_{2}$NiIrO$_{6}$~\cite{La_acta2021,La_PRM2022}, and both are even larger than the atomic upper limit. The outcome of this comparison seems surprising, and it implies the invalidity of the $j_{\rm eff}=1/2$ picture and an additional contribution of the magnetic moments, both of which are discussed below.

We perform hybrid functional+SOC calculations for Y$_{2}$NiIrO$_{6}$ and La$_{2}$NiIrO$_{6}$, and use the $j_{\rm eff}=1/2$ and $j_{\rm eff}=3/2$ basis to project the Ir$^{4+}$ $t_{2g}$$^5$ DOS. As shown in Figs. \ref{fig:7}(a) and \ref{fig:7}(b), the $j_{\rm eff}=1/2$ and $j_{\rm eff}=3/2$ states are severely mixed, and the single $t_{2g}$ hole state of Ir$^{4+}$ ion has nearly equal contributions from $j_{\rm eff}=1/2$ and $j_{\rm eff}=3/2$. Moreover, these calculations show that the Ir$^{4+}$ ions in Y$_{2}$NiIrO$_{6}$ and La$_{2}$NiIrO$_{6}$ each have a local spin/orbital moment of 0.50/0.47 $\mu_{\rm B}$ and 0.48/0.44 $\mu_{\rm B}$, respectively. Both sets of values strongly deviate from the expected spin/orbital moment of 0.33/0.67 $\mu_{\rm B}$ in the $j_{\rm eff}=1/2$ state where the orbital moment is twice as large as the spin moment. As such, we suggest that the $j_{\rm eff}=1/2$ and $j_{\rm eff}=3/2$ states are not good eigenorbitals for the Ir$^{4+}$ ions in Y$_{2}$NiIrO$_{6}$ and La$_{2}$NiIrO$_{6}$. Here the invalidity of the $j_{\rm eff}=1/2$ picture can be ascribed to the $t_{2g}$ splitting by lattice distortion and to Hund exchange. Moreover, a broad band formation is another cause for the invalidity, as the delocalized Ir $5d$ electrons form broad bands in the highly coordinated Ir$^{4+}$ fcc sublattice and they would effectively weaken the SOC effect. To see this, we replace Ni$^{2+}$ by Zn$^{2+}$ and half Ir$^{4+}$ by Ge$^{4+}$ to construct the artificial system Y$_{2}$ZnIr$_{0.5}$Ge$_{0.5}$O$_{6}$ in which the IrZn and ZnGe $ab$-layers alternate along the $c$ axis. Then the Ir-Ir coordination number is reduced from 12 in Y$_{2}$NiIrO$_{6}$ to only 4 in Y$_{2}$ZnIr$_{0.5}$Ge$_{0.5}$O$_{6}$. As shown in Fig. \ref{fig:7}(c), now the mixing of $j_{\rm eff}=1/2$ and $j_{\rm eff}=3/2$ is largely suppressed, and the single $t_{2g}$ hole state is mostly contributed by $j_{\rm eff}=1/2$. Therefore, we conclude that for the highly coordinated Y$_{2}$NiIrO$_{6}$ and La$_{2}$NiIrO$_{6}$ with considerable lattice distortion, the Ir$^{4+}$ ion is not in the $j_{\rm eff}=1/2$ state, but rather in the $S=1/2$ state carrying a finite orbital moment via SOC. In addition, the Ni$^{2+}$ ions in Y$_{2}$NiIrO$_{6}$ (La$_{2}$NiIrO$_{6}$) have each a local spin/orbital moment of 1.68/0.32 $\mu_{\rm B}$ (1.69/0.33 $\mu_{\rm B}$) by the hybrid functional+SOC calculations. As a result, we propose that it is the finite Ni-Ir orbital moments which account for the experimental even larger $\mu_{\rm eff}$\cite{Y_2023,La_acta2021,La_PRM2022} than the theoretical atomic upper limit of the formal Ni$^{2+}$ $S=1$ and Ir$^{4+}$ $S=1/2$ state.

\section{Summary}
In summary, we studied the electronic structure and magnetism of the newly synthesized double perovskites Y$_{2}$NiIrO$_{6}$ and La$_{2}$NiIrO$_{6}$, using the crystal field level diagrams and superexchange pictures, density functional calculations, and Monte Carlo simulations. Our results show that both systems have the Ni$^{2+}$ $t_{2g}$$^{6}e_{g}$$^{2}$ ($S=1$) and Ir$^{4+}$ $t_{2g}$$^{5}$ ($S=1/2$) configurations and are AFM Mott insulators. Moreover, the Ni$^{2+}$-Ir$^{4+}$ AFM coupling turns out to be significantly enhanced from the less distorted La$_{2}$NiIrO$_{6}$ to the more distorted Y$_{2}$NiIrO$_{6}$. Via superexchange model analyses, we addressed that while an ideal cubic structure has a weak FM Ni$^{2+}$-Ir$^{4+}$ coupling, the (stronger) lattice distortion yields (larger) $t_{2g}$-$e_{g}$ orbital mixing which facilitates the (stronger) AFM Ni$^{2+}$-Ir$^{4+}$ coupling. This picture is fully supported by our density functional calculations and Monte Carlo simulations which reproduce the interesting experimental finding that $T_{\rm N}$ is significantly increased from the less distorted La$_{2}$NiIrO$_{6}$ to the more distorted Y$_{2}$NiIrO$_{6}$. We also found that the Ir$^{4+}$ ion is not in the $j_{\rm eff}=1/2$ state due to the $t_{2g}$ splitting by lattice distortion, Hund exchange, and the broad band formation, but it is in the $S=1/2$ state and carries a finite orbital moment via SOC. The Ni$^{2+}$ $S=1$ and Ir$^{4+}$ $S=1/2$ state, in combination with the finite Ni-Ir orbital moment, well accounts for the experimental effective moment. This paper highlights the varying magnetism in Y$_{2}$NiIrO$_{6}$ and La$_{2}$NiIrO$_{6}$ associated with the lattice distortion.

\section*{Acknowledgements}
This work was supported by National Natural Science Foundation
of China (Grants No. 12174062, No. 12241402, and No. 12104307).

\bibliography{rsc}

\begin{thebibliography}{35}%
\makeatletter
\providecommand \@ifxundefined [1]{%
 \@ifx{#1\undefined}
}%
\providecommand \@ifnum [1]{%
 \ifnum #1\expandafter \@firstoftwo
 \else \expandafter \@secondoftwo
 \fi
}%
\providecommand \@ifx [1]{%
 \ifx #1\expandafter \@firstoftwo
 \else \expandafter \@secondoftwo
 \fi
}%
\providecommand \natexlab [1]{#1}%
\providecommand \enquote  [1]{``#1''}%
\providecommand \bibnamefont  [1]{#1}%
\providecommand \bibfnamefont [1]{#1}%
\providecommand \citenamefont [1]{#1}%
\providecommand \href@noop [0]{\@secondoftwo}%
\providecommand \href [0]{\begingroup \@sanitize@url \@href}%
\providecommand \@href[1]{\@@startlink{#1}\@@href}%
\providecommand \@@href[1]{\endgroup#1\@@endlink}%
\providecommand \@sanitize@url [0]{\catcode `\\12\catcode `\$12\catcode
  `\&12\catcode `\#12\catcode `\^12\catcode `\_12\catcode `\%12\relax}%
\providecommand \@@startlink[1]{}%
\providecommand \@@endlink[0]{}%
\providecommand \url  [0]{\begingroup\@sanitize@url \@url }%
\providecommand \@url [1]{\endgroup\@href {#1}{\urlprefix }}%
\providecommand \urlprefix  [0]{URL }%
\providecommand \Eprint [0]{\href }%
\providecommand \doibase [0]{http://dx.doi.org/}%
\providecommand \selectlanguage [0]{\@gobble}%
\providecommand \bibinfo  [0]{\@secondoftwo}%
\providecommand \bibfield  [0]{\@secondoftwo}%
\providecommand \translation [1]{[#1]}%
\providecommand \BibitemOpen [0]{}%
\providecommand \bibitemStop [0]{}%
\providecommand \bibitemNoStop [0]{.\EOS\space}%
\providecommand \EOS [0]{\spacefactor3000\relax}%
\providecommand \BibitemShut  [1]{\csname bibitem#1\endcsname}%
\let\auto@bib@innerbib\@empty
\bibitem [{\citenamefont {Tokura}\ and\ \citenamefont
  {Nagaosa}(2000)}]{Tokura_2000Science}%
  \BibitemOpen
  \bibfield  {author} {\bibinfo {author} {\bibfnamefont {Y.}~\bibnamefont
  {Tokura}}\ and\ \bibinfo {author} {\bibfnamefont {N.}~\bibnamefont
  {Nagaosa}},\ }\bibfield  {title} {\enquote {\bibinfo {title} {Orbital physics
  in transition-metal oxides},}\ }\href {\doibase 10.1126/science.288.5465.462}
  {\bibfield  {journal} {\bibinfo  {journal} {Science}\ }\textbf {\bibinfo
  {volume} {288}},\ \bibinfo {pages} {462--468} (\bibinfo {year}
  {2000})}\BibitemShut {NoStop}%
\bibitem [{\citenamefont {Tokura}(2006)}]{mr}%
  \BibitemOpen
  \bibfield  {author} {\bibinfo {author} {\bibfnamefont {Y.}~\bibnamefont
  {Tokura}},\ }\bibfield  {title} {\enquote {\bibinfo {title} {Critical
  features of colossal magnetoresistive manganites},}\ }\href {\doibase
  10.1088/0034-4885/69/3/R06} {\bibfield  {journal} {\bibinfo  {journal} {Rep.
  Prog. Phys.}\ }\textbf {\bibinfo {volume} {69}},\ \bibinfo {pages} {797}
  (\bibinfo {year} {2006})}\BibitemShut {NoStop}%
\bibitem [{\citenamefont {Kimura}\ \emph {et~al.}(2003)\citenamefont {Kimura},
  \citenamefont {Goto}, \citenamefont {Shintani}, \citenamefont {Ishizaka},
  \citenamefont {Arima},\ and\ \citenamefont {Tokura}}]{ferroelectric}%
  \BibitemOpen
  \bibfield  {author} {\bibinfo {author} {\bibfnamefont {T.}~\bibnamefont
  {Kimura}}, \bibinfo {author} {\bibfnamefont {T.}~\bibnamefont {Goto}},
  \bibinfo {author} {\bibfnamefont {H.}~\bibnamefont {Shintani}}, \bibinfo
  {author} {\bibfnamefont {K.}~\bibnamefont {Ishizaka}}, \bibinfo {author}
  {\bibfnamefont {T.}~\bibnamefont {Arima}}, \ and\ \bibinfo {author}
  {\bibfnamefont {Y.}~\bibnamefont {Tokura}},\ }\bibfield  {title} {\enquote
  {\bibinfo {title} {Magnetic control of ferroelectric polarization},}\ }\href
  {\doibase 10.1038/nature02018} {\bibfield  {journal} {\bibinfo  {journal}
  {Nature}\ }\textbf {\bibinfo {volume} {426}},\ \bibinfo {pages} {55--58}
  (\bibinfo {year} {2003})}\BibitemShut {NoStop}%
\bibitem [{\citenamefont {Khomskii}(2009)}]{multiferroics}%
  \BibitemOpen
  \bibfield  {author} {\bibinfo {author} {\bibfnamefont {D.~I.}\ \bibnamefont
  {Khomskii}},\ }\bibfield  {title} {\enquote {\bibinfo {title} {Classifying
  multiferroics: Mechanisms and effects},}\ }\href
  {https://physics.aps.org/articles/v2/20} {\bibfield  {journal} {\bibinfo
  {journal} {Physics}\ }\textbf {\bibinfo {volume} {2}},\ \bibinfo {pages} {20}
  (\bibinfo {year} {2009})}\BibitemShut {NoStop}%
\bibitem [{\citenamefont {Bednorz}\ and\ \citenamefont
  {M{\"u}ller}(1988)}]{superc}%
  \BibitemOpen
  \bibfield  {author} {\bibinfo {author} {\bibfnamefont {J.~G.}\ \bibnamefont
  {Bednorz}}\ and\ \bibinfo {author} {\bibfnamefont {K.~A.}\ \bibnamefont
  {M{\"u}ller}},\ }\bibfield  {title} {\enquote {\bibinfo {title}
  {Perovskite-type oxides-the new approach to high-${T}_{\mathrm{c}}$
  superconductivity},}\ }\href {\doibase 10.1103/RevModPhys.60.585} {\bibfield
  {journal} {\bibinfo  {journal} {Rev. Mod. Phys.}\ }\textbf {\bibinfo {volume}
  {60}},\ \bibinfo {pages} {585--600} (\bibinfo {year} {1988})}\BibitemShut
  {NoStop}%
\bibitem [{\citenamefont {Vasala}\ and\ \citenamefont
  {Karppinen}(2015)}]{Vasala_2015}%
  \BibitemOpen
  \bibfield  {author} {\bibinfo {author} {\bibfnamefont {Sami}\ \bibnamefont
  {Vasala}}\ and\ \bibinfo {author} {\bibfnamefont {Maarit}\ \bibnamefont
  {Karppinen}},\ }\bibfield  {title} {\enquote {\bibinfo {title}
  {$\mathrm{A}_{2}\mathrm{B}^{\prime}\mathrm{B}^{\prime\prime}\mathrm{O}_{6}$
  perovskites: A review},}\ }\href {\doibase
  https://doi.org/10.1016/j.progsolidstchem.2014.08.001} {\bibfield  {journal}
  {\bibinfo  {journal} {Prog. Solid State Chem.}\ }\textbf {\bibinfo {volume}
  {43}},\ \bibinfo {pages} {1--36} (\bibinfo {year} {2015})}\BibitemShut
  {NoStop}%
\bibitem [{\citenamefont {Kato}\ \emph {et~al.}(2002)\citenamefont {Kato},
  \citenamefont {Okuda}, \citenamefont {Okimoto}, \citenamefont {Tomioka},
  \citenamefont {Takenoya}, \citenamefont {Ohkubo}, \citenamefont {Kawasaki},\
  and\ \citenamefont {Tokura}}]{Sr2CrReO6}%
  \BibitemOpen
  \bibfield  {author} {\bibinfo {author} {\bibfnamefont {H.}~\bibnamefont
  {Kato}}, \bibinfo {author} {\bibfnamefont {T.}~\bibnamefont {Okuda}},
  \bibinfo {author} {\bibfnamefont {Y.}~\bibnamefont {Okimoto}}, \bibinfo
  {author} {\bibfnamefont {Y.}~\bibnamefont {Tomioka}}, \bibinfo {author}
  {\bibfnamefont {Y.}~\bibnamefont {Takenoya}}, \bibinfo {author}
  {\bibfnamefont {A.}~\bibnamefont {Ohkubo}}, \bibinfo {author} {\bibfnamefont
  {M.}~\bibnamefont {Kawasaki}}, \ and\ \bibinfo {author} {\bibfnamefont
  {Y.}~\bibnamefont {Tokura}},\ }\bibfield  {title} {\enquote {\bibinfo {title}
  {Metallic ordered double-perovskite
  $\mathrm{S}{\mathrm{r}}_{2}\mathrm{C}{\mathrm{r}}\mathrm{R}{\mathrm{e}}\mathrm{O}_{6}$
  with maximal curie temperature of 635 $\mathrm{K}$},}\ }\href {\doibase
  10.1063/1.1493646} {\bibfield  {journal} {\bibinfo  {journal} {Appl. Phys.
  Lett.}\ }\textbf {\bibinfo {volume} {81}},\ \bibinfo {pages} {328--330}
  (\bibinfo {year} {2002})}\BibitemShut {NoStop}%
\bibitem [{\citenamefont {Krockenberger}\ \emph {et~al.}(2007)\citenamefont
  {Krockenberger}, \citenamefont {Mogare}, \citenamefont {Reehuis},
  \citenamefont {Tovar}, \citenamefont {Jansen}, \citenamefont {Vaitheeswaran},
  \citenamefont {Kanchana}, \citenamefont {Bultmark}, \citenamefont {Delin},
  \citenamefont {Wilhelm}, \citenamefont {Rogalev}, \citenamefont {Winkler},\
  and\ \citenamefont {Alff}}]{Sr2CrOsO6}%
  \BibitemOpen
  \bibfield  {author} {\bibinfo {author} {\bibfnamefont {Y.}~\bibnamefont
  {Krockenberger}}, \bibinfo {author} {\bibfnamefont {K.}~\bibnamefont
  {Mogare}}, \bibinfo {author} {\bibfnamefont {M.}~\bibnamefont {Reehuis}},
  \bibinfo {author} {\bibfnamefont {M.}~\bibnamefont {Tovar}}, \bibinfo
  {author} {\bibfnamefont {M.}~\bibnamefont {Jansen}}, \bibinfo {author}
  {\bibfnamefont {G.}~\bibnamefont {Vaitheeswaran}}, \bibinfo {author}
  {\bibfnamefont {V.}~\bibnamefont {Kanchana}}, \bibinfo {author}
  {\bibfnamefont {F.}~\bibnamefont {Bultmark}}, \bibinfo {author}
  {\bibfnamefont {A.}~\bibnamefont {Delin}}, \bibinfo {author} {\bibfnamefont
  {F.}~\bibnamefont {Wilhelm}}, \bibinfo {author} {\bibfnamefont
  {A.}~\bibnamefont {Rogalev}}, \bibinfo {author} {\bibfnamefont
  {A.}~\bibnamefont {Winkler}}, \ and\ \bibinfo {author} {\bibfnamefont
  {L.}~\bibnamefont {Alff}},\ }\bibfield  {title} {\enquote {\bibinfo {title}
  {$\mathrm{S}{\mathrm{r}}_{2}\mathrm{C}{\mathrm{r}}\mathrm{O}{\mathrm{s}}\mathrm{O}_{6}$:
  End point of a spin-polarized metal-insulator transition by $5d$ band
  filling},}\ }\href {\doibase 10.1103/PhysRevB.75.020404} {\bibfield
  {journal} {\bibinfo  {journal} {Phys. Rev. B}\ }\textbf {\bibinfo {volume}
  {75}},\ \bibinfo {pages} {020404(R)} (\bibinfo {year} {2007})}\BibitemShut
  {NoStop}%
\bibitem [{\citenamefont {Morrow}\ \emph {et~al.}(2016)\citenamefont {Morrow},
  \citenamefont {Soliz}, \citenamefont {Hauser}, \citenamefont {Gallagher},
  \citenamefont {Susner}, \citenamefont {Sumption}, \citenamefont {Aczel},
  \citenamefont {Yan}, \citenamefont {Yang},\ and\ \citenamefont
  {Woodward}}]{Ca2CrOsO6}%
  \BibitemOpen
  \bibfield  {author} {\bibinfo {author} {\bibfnamefont {R.}~\bibnamefont
  {Morrow}}, \bibinfo {author} {\bibfnamefont {J.~R.}\ \bibnamefont {Soliz}},
  \bibinfo {author} {\bibfnamefont {A.~J.}\ \bibnamefont {Hauser}}, \bibinfo
  {author} {\bibfnamefont {J.~C.}\ \bibnamefont {Gallagher}}, \bibinfo {author}
  {\bibfnamefont {M.~A.}\ \bibnamefont {Susner}}, \bibinfo {author}
  {\bibfnamefont {M.~D.}\ \bibnamefont {Sumption}}, \bibinfo {author}
  {\bibfnamefont {A.~A.}\ \bibnamefont {Aczel}}, \bibinfo {author}
  {\bibfnamefont {J.}~\bibnamefont {Yan}}, \bibinfo {author} {\bibfnamefont
  {F.}~\bibnamefont {Yang}}, \ and\ \bibinfo {author} {\bibfnamefont {P.~M.}\
  \bibnamefont {Woodward}},\ }\bibfield  {title} {\enquote {\bibinfo {title}
  {The effect of chemical pressure on the structure and properties of
  $\mathrm{A}_{2}\mathrm{C}{\mathrm{r}}\mathrm{O}{\mathrm{s}}\mathrm{O}_{6}$
  ($\mathrm{A}=\mathrm{S}{\mathrm{r}},\mathrm{C}{\mathrm{a}}$) ferrimagnetic
  double perovskite},}\ }\href {\doibase
  https://doi.org/10.1016/j.jssc.2016.02.025} {\bibfield  {journal} {\bibinfo
  {journal} {J. Solid State Chem.}\ }\textbf {\bibinfo {volume} {238}},\
  \bibinfo {pages} {46--52} (\bibinfo {year} {2016})}\BibitemShut {NoStop}%
\bibitem [{\citenamefont {Cao}\ \emph {et~al.}(2014)\citenamefont {Cao},
  \citenamefont {Qi}, \citenamefont {Li}, \citenamefont {Terzic}, \citenamefont
  {Yuan}, \citenamefont {DeLong}, \citenamefont {Murthy},\ and\ \citenamefont
  {Kaul}}]{Sr2YIrO6}%
  \BibitemOpen
  \bibfield  {author} {\bibinfo {author} {\bibfnamefont {G.}~\bibnamefont
  {Cao}}, \bibinfo {author} {\bibfnamefont {T.~F.}\ \bibnamefont {Qi}},
  \bibinfo {author} {\bibfnamefont {L.}~\bibnamefont {Li}}, \bibinfo {author}
  {\bibfnamefont {J.}~\bibnamefont {Terzic}}, \bibinfo {author} {\bibfnamefont
  {S.~J.}\ \bibnamefont {Yuan}}, \bibinfo {author} {\bibfnamefont {L.~E.}\
  \bibnamefont {DeLong}}, \bibinfo {author} {\bibfnamefont {G.}~\bibnamefont
  {Murthy}}, \ and\ \bibinfo {author} {\bibfnamefont {R.~K.}\ \bibnamefont
  {Kaul}},\ }\bibfield  {title} {\enquote {\bibinfo {title} {Novel magnetism of
  $\mathrm{I}{\mathrm{r}}^{5+}(5{d}^{4})$ ions in the double perovskite
  $\mathrm{S}{\mathrm{r}}_{2}\mathrm{Y}\mathrm{I}{\mathrm{r}}\mathrm{O}_{6}$},}\
  }\href {\doibase 10.1103/PhysRevLett.112.056402} {\bibfield  {journal}
  {\bibinfo  {journal} {Phys. Rev. Lett.}\ }\textbf {\bibinfo {volume} {112}},\
  \bibinfo {pages} {056402} (\bibinfo {year} {2014})}\BibitemShut {NoStop}%
\bibitem [{\citenamefont {Dey}\ \emph {et~al.}(2016)\citenamefont {Dey},
  \citenamefont {Maljuk}, \citenamefont {Efremov}, \citenamefont {Kataeva},
  \citenamefont {Gass}, \citenamefont {Blum}, \citenamefont {Steckel},
  \citenamefont {Gruner}, \citenamefont {Ritschel}, \citenamefont {Wolter},
  \citenamefont {Geck}, \citenamefont {Hess}, \citenamefont {Koepernik},
  \citenamefont {van~den Brink}, \citenamefont {Wurmehl},\ and\ \citenamefont
  {B\"uchner}}]{Ba2YIrO6_2016}%
  \BibitemOpen
  \bibfield  {author} {\bibinfo {author} {\bibfnamefont {T.}~\bibnamefont
  {Dey}}, \bibinfo {author} {\bibfnamefont {A.}~\bibnamefont {Maljuk}},
  \bibinfo {author} {\bibfnamefont {D.~V.}\ \bibnamefont {Efremov}}, \bibinfo
  {author} {\bibfnamefont {O.}~\bibnamefont {Kataeva}}, \bibinfo {author}
  {\bibfnamefont {S.}~\bibnamefont {Gass}}, \bibinfo {author} {\bibfnamefont
  {C.~G.~F.}\ \bibnamefont {Blum}}, \bibinfo {author} {\bibfnamefont
  {F.}~\bibnamefont {Steckel}}, \bibinfo {author} {\bibfnamefont
  {D.}~\bibnamefont {Gruner}}, \bibinfo {author} {\bibfnamefont
  {T.}~\bibnamefont {Ritschel}}, \bibinfo {author} {\bibfnamefont {A.~U.~B.}\
  \bibnamefont {Wolter}}, \bibinfo {author} {\bibfnamefont {J.}~\bibnamefont
  {Geck}}, \bibinfo {author} {\bibfnamefont {C.}~\bibnamefont {Hess}}, \bibinfo
  {author} {\bibfnamefont {K.}~\bibnamefont {Koepernik}}, \bibinfo {author}
  {\bibfnamefont {J.}~\bibnamefont {van~den Brink}}, \bibinfo {author}
  {\bibfnamefont {S.}~\bibnamefont {Wurmehl}}, \ and\ \bibinfo {author}
  {\bibfnamefont {B.}~\bibnamefont {B\"uchner}},\ }\bibfield  {title} {\enquote
  {\bibinfo {title}
  {$\mathrm{B}{\mathrm{a}}_{2}\mathrm{Y}\mathrm{I}{\mathrm{r}}\mathrm{O}_{6}$:
  A cubic double perovskite material with $\mathrm{I}{\mathrm{r}}^{5+}$
  ions},}\ }\href {\doibase 10.1103/PhysRevB.93.014434} {\bibfield  {journal}
  {\bibinfo  {journal} {Phys. Rev. B}\ }\textbf {\bibinfo {volume} {93}},\
  \bibinfo {pages} {014434} (\bibinfo {year} {2016})}\BibitemShut {NoStop}%
\bibitem [{\citenamefont {Terzic}\ \emph {et~al.}(2017)\citenamefont {Terzic},
  \citenamefont {Zheng}, \citenamefont {Ye}, \citenamefont {Zhao},
  \citenamefont {Schlottmann}, \citenamefont {De~Long}, \citenamefont {Yuan},\
  and\ \citenamefont {Cao}}]{Ba2YIrO6_2017}%
  \BibitemOpen
  \bibfield  {author} {\bibinfo {author} {\bibfnamefont {J.}~\bibnamefont
  {Terzic}}, \bibinfo {author} {\bibfnamefont {H.}~\bibnamefont {Zheng}},
  \bibinfo {author} {\bibfnamefont {Feng}\ \bibnamefont {Ye}}, \bibinfo
  {author} {\bibfnamefont {H.~D.}\ \bibnamefont {Zhao}}, \bibinfo {author}
  {\bibfnamefont {P.}~\bibnamefont {Schlottmann}}, \bibinfo {author}
  {\bibfnamefont {L.~E.}\ \bibnamefont {De~Long}}, \bibinfo {author}
  {\bibfnamefont {S.~J.}\ \bibnamefont {Yuan}}, \ and\ \bibinfo {author}
  {\bibfnamefont {G.}~\bibnamefont {Cao}},\ }\bibfield  {title} {\enquote
  {\bibinfo {title} {Evidence for a low-temperature magnetic ground state in
  double-perovskite iridates with $\mathrm{I}{\mathrm{r}}^{5+}(5{d}^{4})$
  ions},}\ }\href {\doibase 10.1103/PhysRevB.96.064436} {\bibfield  {journal}
  {\bibinfo  {journal} {Phys. Rev. B}\ }\textbf {\bibinfo {volume} {96}},\
  \bibinfo {pages} {064436} (\bibinfo {year} {2017})}\BibitemShut {NoStop}%
\bibitem [{\citenamefont {Deng}\ \emph {et~al.}(2023)\citenamefont {Deng},
  \citenamefont {Wang}, \citenamefont {Wang}, \citenamefont {Shen},
  \citenamefont {Zhang}, \citenamefont {Chen}, \citenamefont {Feng},
  \citenamefont {Xu}, \citenamefont {Peng}, \citenamefont {Li}, \citenamefont
  {Zhao}, \citenamefont {Wang}, \citenamefont {Valvidares}, \citenamefont
  {Francoual}, \citenamefont {Leupold}, \citenamefont {Hu}, \citenamefont
  {Tjeng}, \citenamefont {Li}, \citenamefont {Croft}, \citenamefont {Zhang},
  \citenamefont {Liu}, \citenamefont {He}, \citenamefont {Hu}, \citenamefont
  {Sun}, \citenamefont {Greenblatt},\ and\ \citenamefont {Jin}}]{Y_2023}%
  \BibitemOpen
  \bibfield  {author} {\bibinfo {author} {\bibfnamefont {Z.}~\bibnamefont
  {Deng}}, \bibinfo {author} {\bibfnamefont {X.}~\bibnamefont {Wang}}, \bibinfo
  {author} {\bibfnamefont {M.}~\bibnamefont {Wang}}, \bibinfo {author}
  {\bibfnamefont {F.}~\bibnamefont {Shen}}, \bibinfo {author} {\bibfnamefont
  {J.}~\bibnamefont {Zhang}}, \bibinfo {author} {\bibfnamefont
  {Y.}~\bibnamefont {Chen}}, \bibinfo {author} {\bibfnamefont {H.~L.}\
  \bibnamefont {Feng}}, \bibinfo {author} {\bibfnamefont {J.}~\bibnamefont
  {Xu}}, \bibinfo {author} {\bibfnamefont {Y.}~\bibnamefont {Peng}}, \bibinfo
  {author} {\bibfnamefont {W.}~\bibnamefont {Li}}, \bibinfo {author}
  {\bibfnamefont {J.}~\bibnamefont {Zhao}}, \bibinfo {author} {\bibfnamefont
  {X.}~\bibnamefont {Wang}}, \bibinfo {author} {\bibfnamefont {M.}~\bibnamefont
  {Valvidares}}, \bibinfo {author} {\bibfnamefont {S.}~\bibnamefont
  {Francoual}}, \bibinfo {author} {\bibfnamefont {O.}~\bibnamefont {Leupold}},
  \bibinfo {author} {\bibfnamefont {Z.}~\bibnamefont {Hu}}, \bibinfo {author}
  {\bibfnamefont {L.~H.}\ \bibnamefont {Tjeng}}, \bibinfo {author}
  {\bibfnamefont {M.~R.}\ \bibnamefont {Li}}, \bibinfo {author} {\bibfnamefont
  {M.}~\bibnamefont {Croft}}, \bibinfo {author} {\bibfnamefont
  {Y.}~\bibnamefont {Zhang}}, \bibinfo {author} {\bibfnamefont
  {E.}~\bibnamefont {Liu}}, \bibinfo {author} {\bibfnamefont {L.}~\bibnamefont
  {He}}, \bibinfo {author} {\bibfnamefont {F.}~\bibnamefont {Hu}}, \bibinfo
  {author} {\bibfnamefont {J.}~\bibnamefont {Sun}}, \bibinfo {author}
  {\bibfnamefont {M.}~\bibnamefont {Greenblatt}}, \ and\ \bibinfo {author}
  {\bibfnamefont {C.}~\bibnamefont {Jin}},\ }\bibfield  {title} {\enquote
  {\bibinfo {title} {Giant exchange-bias-like effect at low cooling fields
  induced by pinned magnetic domains in
  $\mathrm{Y}_{2}\mathrm{N}{\mathrm{i}}\mathrm{I}{\mathrm{r}}\mathrm{O}_{6}$
  double perovskite},}\ }\href {\doibase 10.1002/adma.202209759} {\bibfield
  {journal} {\bibinfo  {journal} {Adv. Mater.}\ ,\ \bibinfo {pages} {2209759}}
  (\bibinfo {year} {2023})}\BibitemShut {NoStop}%
\bibitem [{\citenamefont {Kayser}\ \emph {et~al.}(2021)\citenamefont {Kayser},
  \citenamefont {Mu{\~n}oz}, \citenamefont {Mart{\'\i}nez}, \citenamefont
  {Fauth}, \citenamefont {Fern{\'a}ndez-D{\'\i}az},\ and\ \citenamefont
  {Alonso}}]{La_acta2021}%
  \BibitemOpen
  \bibfield  {author} {\bibinfo {author} {\bibfnamefont {P.}~\bibnamefont
  {Kayser}}, \bibinfo {author} {\bibfnamefont {A.}~\bibnamefont {Mu{\~n}oz}},
  \bibinfo {author} {\bibfnamefont {J.L.}\ \bibnamefont {Mart{\'\i}nez}},
  \bibinfo {author} {\bibfnamefont {F.}~\bibnamefont {Fauth}}, \bibinfo
  {author} {\bibfnamefont {M.T.}\ \bibnamefont {Fern{\'a}ndez-D{\'\i}az}}, \
  and\ \bibinfo {author} {\bibfnamefont {J.A.}\ \bibnamefont {Alonso}},\
  }\bibfield  {title} {\enquote {\bibinfo {title} {Enhancing the
  $\mathrm{N}${\'e}el temperature in 3$d$/5$d$
  ${R}_{2}\mathrm{N}{\mathrm{i}}\mathrm{I}{\mathrm{r}}\mathrm{O}_{6}$
  $({R}=\mathrm{L}{\mathrm{a}}, \mathrm{P}{\mathrm{r}},
  \mathrm{N}{\mathrm{d}})$ double perovskites by reducing the ${R}^{3+}$ ionic
  radii},}\ }\href {\doibase https://doi.org/10.1016/j.actamat.2021.116684}
  {\bibfield  {journal} {\bibinfo  {journal} {Acta Mater.}\ }\textbf {\bibinfo
  {volume} {207}},\ \bibinfo {pages} {116684} (\bibinfo {year}
  {2021})}\BibitemShut {NoStop}%
\bibitem [{\citenamefont {Ferreira}\ \emph {et~al.}(2021)\citenamefont
  {Ferreira}, \citenamefont {Calder}, \citenamefont {Parker}, \citenamefont
  {Upton}, \citenamefont {Sefat},\ and\ \citenamefont {Loye}}]{La_PRM2021}%
  \BibitemOpen
  \bibfield  {author} {\bibinfo {author} {\bibfnamefont {T.}~\bibnamefont
  {Ferreira}}, \bibinfo {author} {\bibfnamefont {S.}~\bibnamefont {Calder}},
  \bibinfo {author} {\bibfnamefont {D.~S.}\ \bibnamefont {Parker}}, \bibinfo
  {author} {\bibfnamefont {M.~H.}\ \bibnamefont {Upton}}, \bibinfo {author}
  {\bibfnamefont {A.~S.}\ \bibnamefont {Sefat}}, \ and\ \bibinfo {author}
  {\bibfnamefont {H.-C.~zur}\ \bibnamefont {Loye}},\ }\bibfield  {title}
  {\enquote {\bibinfo {title} {Relationship between a-site cation and magnetic
  structure in $3d\text{\ensuremath{-}}5d\text{\ensuremath{-}}4f$ double
  perovskite iridates
  ${Ln}\mathrm{N}{\mathrm{i}}\mathrm{I}{\mathrm{r}}\mathrm{O}_{6}$
  $({Ln}=\mathrm{L}{\mathrm{a}}, \mathrm{P}{\mathrm{r}},
  \mathrm{N}{\mathrm{d}})$},}\ }\href {\doibase
  10.1103/PhysRevMaterials.5.064408} {\bibfield  {journal} {\bibinfo  {journal}
  {Phys. Rev. Mater.}\ }\textbf {\bibinfo {volume} {5}},\ \bibinfo {pages}
  {064408} (\bibinfo {year} {2021})}\BibitemShut {NoStop}%
\bibitem [{\citenamefont {Sharma}\ \emph {et~al.}(2022)\citenamefont {Sharma},
  \citenamefont {Ritter}, \citenamefont {Adroja}, \citenamefont {Stenning},
  \citenamefont {Sundaresan},\ and\ \citenamefont {Langridge}}]{La_PRM2022}%
  \BibitemOpen
  \bibfield  {author} {\bibinfo {author} {\bibfnamefont {S.}~\bibnamefont
  {Sharma}}, \bibinfo {author} {\bibfnamefont {C.}~\bibnamefont {Ritter}},
  \bibinfo {author} {\bibfnamefont {D.~T.}\ \bibnamefont {Adroja}}, \bibinfo
  {author} {\bibfnamefont {G.~B.}\ \bibnamefont {Stenning}}, \bibinfo {author}
  {\bibfnamefont {A.}~\bibnamefont {Sundaresan}}, \ and\ \bibinfo {author}
  {\bibfnamefont {S.}~\bibnamefont {Langridge}},\ }\bibfield  {title} {\enquote
  {\bibinfo {title} {Magnetic structure of the double perovskite
  $\mathrm{L}{\mathrm{a}}_{2}\mathrm{N}{\mathrm{i}}\mathrm{I}{\mathrm{r}}\mathrm{O}_{6}$
  investigated using neutron diffraction},}\ }\href {\doibase
  10.1103/PhysRevMaterials.6.014407} {\bibfield  {journal} {\bibinfo  {journal}
  {Phys. Rev. Mater.}\ }\textbf {\bibinfo {volume} {6}},\ \bibinfo {pages}
  {014407} (\bibinfo {year} {2022})}\BibitemShut {NoStop}%
\bibitem [{\citenamefont {Khomskii}(2014)}]{Khomskii_2001}%
  \BibitemOpen
  \bibfield  {author} {\bibinfo {author} {\bibfnamefont {D.~I.}\ \bibnamefont
  {Khomskii}},\ }\href@noop {} {\emph {\bibinfo {title} {Transition Metal
  Compounds}}}\ (\bibinfo  {publisher} {Cambridge University Press, Cambridge,
  UK},\ \bibinfo {year} {2014})\BibitemShut {NoStop}%
\bibitem [{\citenamefont {Jin}\ \emph {et~al.}(2022)\citenamefont {Jin},
  \citenamefont {Chun}, \citenamefont {Kim}, \citenamefont {Casa},
  \citenamefont {Ruff}, \citenamefont {Won}, \citenamefont {Lee}, \citenamefont
  {Hur},\ and\ \citenamefont {Kim}}]{La_PRB2022}%
  \BibitemOpen
  \bibfield  {author} {\bibinfo {author} {\bibfnamefont {W.}~\bibnamefont
  {Jin}}, \bibinfo {author} {\bibfnamefont {S.~H.}\ \bibnamefont {Chun}},
  \bibinfo {author} {\bibfnamefont {J.}~\bibnamefont {Kim}}, \bibinfo {author}
  {\bibfnamefont {D.}~\bibnamefont {Casa}}, \bibinfo {author} {\bibfnamefont
  {J.~P.~C.}\ \bibnamefont {Ruff}}, \bibinfo {author} {\bibfnamefont {C.~J.}\
  \bibnamefont {Won}}, \bibinfo {author} {\bibfnamefont {K.~D.}\ \bibnamefont
  {Lee}}, \bibinfo {author} {\bibfnamefont {N.}~\bibnamefont {Hur}}, \ and\
  \bibinfo {author} {\bibfnamefont {Y.-J.}\ \bibnamefont {Kim}},\ }\bibfield
  {title} {\enquote {\bibinfo {title} {Magnetic excitations in the
  double-perovskite iridates
  $\mathrm{L}{\mathrm{a}}_{2}{M}\mathrm{I}{\mathrm{r}}\mathrm{O}_{6}$
  $({M}=\mathrm{C}{\mathrm{o}}, \mathrm{N}{\mathrm{i}},
  \mathrm{Z}{\mathrm{n}})$ mediated by $3d\text{\ensuremath{-}}5d$
  hybridization},}\ }\href {\doibase 10.1103/PhysRevB.105.054419} {\bibfield
  {journal} {\bibinfo  {journal} {Phys. Rev. B}\ }\textbf {\bibinfo {volume}
  {105}},\ \bibinfo {pages} {054419} (\bibinfo {year} {2022})}\BibitemShut
  {NoStop}%
\bibitem [{\citenamefont {Blaha}\ \emph {et~al.}(2020)\citenamefont {Blaha},
  \citenamefont {Schwarz}, \citenamefont {Tran}, \citenamefont {Laskowski},
  \citenamefont {Madsen},\ and\ \citenamefont {Marks}}]{WIEN2k}%
  \BibitemOpen
  \bibfield  {author} {\bibinfo {author} {\bibfnamefont {P.}~\bibnamefont
  {Blaha}}, \bibinfo {author} {\bibfnamefont {K.}~\bibnamefont {Schwarz}},
  \bibinfo {author} {\bibfnamefont {F.}~\bibnamefont {Tran}}, \bibinfo {author}
  {\bibfnamefont {R.}~\bibnamefont {Laskowski}}, \bibinfo {author}
  {\bibfnamefont {G.~K.~H.}\ \bibnamefont {Madsen}}, \ and\ \bibinfo {author}
  {\bibfnamefont {L.~D.}\ \bibnamefont {Marks}},\ }\bibfield  {title} {\enquote
  {\bibinfo {title} {Wien2k: An $\mathrm{A}\mathrm{P}\mathrm{W}$+lo program for
  calculating the properties of solids},}\ }\href
  {https://doi.org/10.1063/1.5143061} {\bibfield  {journal} {\bibinfo
  {journal} {J. Chem. Phys.}\ }\textbf {\bibinfo {volume} {152}},\ \bibinfo
  {pages} {074101} (\bibinfo {year} {2020})}\BibitemShut {NoStop}%
\bibitem [{\citenamefont {Becke}(1993)}]{Becke_1993}%
  \BibitemOpen
  \bibfield  {author} {\bibinfo {author} {\bibfnamefont {A.~D.}\ \bibnamefont
  {Becke}},\ }\bibfield  {title} {\enquote {\bibinfo {title} {{A new mixing of
  Hartree–Fock and local density‐functional theories}},}\ }\href {\doibase
  10.1063/1.464304} {\bibfield  {journal} {\bibinfo  {journal} {J. Chem.
  Phys.}\ }\textbf {\bibinfo {volume} {98}},\ \bibinfo {pages} {1372--1377}
  (\bibinfo {year} {1993})}\BibitemShut {NoStop}%
\bibitem [{\citenamefont {Becke}(1996)}]{Becke_1996}%
  \BibitemOpen
  \bibfield  {author} {\bibinfo {author} {\bibfnamefont {A.~D.}\ \bibnamefont
  {Becke}},\ }\bibfield  {title} {\enquote {\bibinfo {title}
  {{Density‐functional thermochemistry. IV. A new dynamical correlation
  functional and implications for exact‐exchange mixing}},}\ }\href {\doibase
  10.1063/1.470829} {\bibfield  {journal} {\bibinfo  {journal} {J. Chem.
  Phys.}\ }\textbf {\bibinfo {volume} {104}},\ \bibinfo {pages} {1040--1046}
  (\bibinfo {year} {1996})}\BibitemShut {NoStop}%
\bibitem [{\citenamefont {Perdew}\ \emph {et~al.}(1996)\citenamefont {Perdew},
  \citenamefont {Ernzerhof},\ and\ \citenamefont {Burke}}]{Fock25}%
  \BibitemOpen
  \bibfield  {author} {\bibinfo {author} {\bibfnamefont {J.~P.}\ \bibnamefont
  {Perdew}}, \bibinfo {author} {\bibfnamefont {M.}~\bibnamefont {Ernzerhof}}, \
  and\ \bibinfo {author} {\bibfnamefont {K.}~\bibnamefont {Burke}},\ }\bibfield
   {title} {\enquote {\bibinfo {title} {{Rationale for mixing exact exchange
  with density functional approximations}},}\ }\href {\doibase
  10.1063/1.472933} {\bibfield  {journal} {\bibinfo  {journal} {J. Chem.
  Phys.}\ }\textbf {\bibinfo {volume} {105}},\ \bibinfo {pages} {9982--9985}
  (\bibinfo {year} {1996})}\BibitemShut {NoStop}%
\bibitem [{\citenamefont {Tran}\ \emph {et~al.}(2006)\citenamefont {Tran},
  \citenamefont {Blaha}, \citenamefont {Schwarz},\ and\ \citenamefont
  {Nov\'ak}}]{hf_1}%
  \BibitemOpen
  \bibfield  {author} {\bibinfo {author} {\bibfnamefont {F.}~\bibnamefont
  {Tran}}, \bibinfo {author} {\bibfnamefont {P.}~\bibnamefont {Blaha}},
  \bibinfo {author} {\bibfnamefont {K.}~\bibnamefont {Schwarz}}, \ and\
  \bibinfo {author} {\bibfnamefont {P.}~\bibnamefont {Nov\'ak}},\ }\bibfield
  {title} {\enquote {\bibinfo {title} {Hybrid exchange-correlation energy
  functionals for strongly correlated electrons: Applications to
  transition-metal monoxides},}\ }\href {\doibase 10.1103/PhysRevB.74.155108}
  {\bibfield  {journal} {\bibinfo  {journal} {Phys. Rev. B}\ }\textbf {\bibinfo
  {volume} {74}},\ \bibinfo {pages} {155108} (\bibinfo {year}
  {2006})}\BibitemShut {NoStop}%
\bibitem [{\citenamefont {Nov{\'a}k}\ \emph {et~al.}(2006)\citenamefont
  {Nov{\'a}k}, \citenamefont {Kune{\v s}}, \citenamefont {Chaput},\ and\
  \citenamefont {Pickett}}]{hf_2}%
  \BibitemOpen
  \bibfield  {author} {\bibinfo {author} {\bibfnamefont {P.}~\bibnamefont
  {Nov{\'a}k}}, \bibinfo {author} {\bibfnamefont {J.}~\bibnamefont {Kune{\v
  s}}}, \bibinfo {author} {\bibfnamefont {L.}~\bibnamefont {Chaput}}, \ and\
  \bibinfo {author} {\bibfnamefont {W.~E.}\ \bibnamefont {Pickett}},\
  }\bibfield  {title} {\enquote {\bibinfo {title} {Exact exchange for
  correlated electrons},}\ }\href
  {https://onlinelibrary.wiley.com/doi/abs/10.1002/pssb.200541371} {\bibfield
  {journal} {\bibinfo  {journal} {Phys. Status Solidi B}\ }\textbf {\bibinfo
  {volume} {243}},\ \bibinfo {pages} {563--572} (\bibinfo {year}
  {2006})}\BibitemShut {NoStop}%
\bibitem [{\citenamefont {Anisimov}\ \emph {et~al.}(1993)\citenamefont
  {Anisimov}, \citenamefont {Solovyev}, \citenamefont {Korotin}, \citenamefont
  {Czy\ifmmode~\dot{z}\else \.{z}\fi{}yk},\ and\ \citenamefont
  {Sawatzky}}]{Anisimov_1993}%
  \BibitemOpen
  \bibfield  {author} {\bibinfo {author} {\bibfnamefont {V.~I.}\ \bibnamefont
  {Anisimov}}, \bibinfo {author} {\bibfnamefont {I.~V.}\ \bibnamefont
  {Solovyev}}, \bibinfo {author} {\bibfnamefont {M.~A.}\ \bibnamefont
  {Korotin}}, \bibinfo {author} {\bibfnamefont {M.~T.}\ \bibnamefont
  {Czy\ifmmode~\dot{z}\else \.{z}\fi{}yk}}, \ and\ \bibinfo {author}
  {\bibfnamefont {G.~A.}\ \bibnamefont {Sawatzky}},\ }\bibfield  {title}
  {\enquote {\bibinfo {title} {Density-functional theory and nio photoemission
  spectra},}\ }\href {https://link.aps.org/doi/10.1103/PhysRevB.48.16929}
  {\bibfield  {journal} {\bibinfo  {journal} {Phys. Rev. B}\ }\textbf {\bibinfo
  {volume} {48}},\ \bibinfo {pages} {16929} (\bibinfo {year}
  {1993})}\BibitemShut {NoStop}%
\bibitem [{\citenamefont {Yuan}\ \emph {et~al.}(2017)\citenamefont {Yuan},
  \citenamefont {Clancy}, \citenamefont {Cook}, \citenamefont {Thompson},
  \citenamefont {Greedan}, \citenamefont {Cao}, \citenamefont {Jeon},
  \citenamefont {Noh}, \citenamefont {Upton}, \citenamefont {Casa},
  \citenamefont {Gog}, \citenamefont {Paramekanti},\ and\ \citenamefont
  {Kim}}]{JH_1}%
  \BibitemOpen
  \bibfield  {author} {\bibinfo {author} {\bibfnamefont {B.}~\bibnamefont
  {Yuan}}, \bibinfo {author} {\bibfnamefont {J.~P.}\ \bibnamefont {Clancy}},
  \bibinfo {author} {\bibfnamefont {A.~M.}\ \bibnamefont {Cook}}, \bibinfo
  {author} {\bibfnamefont {C.~M.}\ \bibnamefont {Thompson}}, \bibinfo {author}
  {\bibfnamefont {J.}~\bibnamefont {Greedan}}, \bibinfo {author} {\bibfnamefont
  {G.}~\bibnamefont {Cao}}, \bibinfo {author} {\bibfnamefont {B.~C.}\
  \bibnamefont {Jeon}}, \bibinfo {author} {\bibfnamefont {T.~W.}\ \bibnamefont
  {Noh}}, \bibinfo {author} {\bibfnamefont {M.~H.}\ \bibnamefont {Upton}},
  \bibinfo {author} {\bibfnamefont {D.}~\bibnamefont {Casa}}, \bibinfo {author}
  {\bibfnamefont {T.}~\bibnamefont {Gog}}, \bibinfo {author} {\bibfnamefont
  {A.}~\bibnamefont {Paramekanti}}, \ and\ \bibinfo {author} {\bibfnamefont
  {Y.-J.}\ \bibnamefont {Kim}},\ }\bibfield  {title} {\enquote {\bibinfo
  {title} {Determination of hund's coupling in $5d$ oxides using resonant
  inelastic x-ray scattering},}\ }\href {\doibase 10.1103/PhysRevB.95.235114}
  {\bibfield  {journal} {\bibinfo  {journal} {Phys. Rev. B}\ }\textbf {\bibinfo
  {volume} {95}},\ \bibinfo {pages} {235114} (\bibinfo {year}
  {2017})}\BibitemShut {NoStop}%
\bibitem [{\citenamefont {Paramekanti}\ \emph {et~al.}(2018)\citenamefont
  {Paramekanti}, \citenamefont {Singh}, \citenamefont {Yuan}, \citenamefont
  {Casa}, \citenamefont {Said}, \citenamefont {Kim},\ and\ \citenamefont
  {Christianson}}]{JH_2}%
  \BibitemOpen
  \bibfield  {author} {\bibinfo {author} {\bibfnamefont {A.}~\bibnamefont
  {Paramekanti}}, \bibinfo {author} {\bibfnamefont {D.~J.}\ \bibnamefont
  {Singh}}, \bibinfo {author} {\bibfnamefont {B.}~\bibnamefont {Yuan}},
  \bibinfo {author} {\bibfnamefont {D.}~\bibnamefont {Casa}}, \bibinfo {author}
  {\bibfnamefont {A.}~\bibnamefont {Said}}, \bibinfo {author} {\bibfnamefont
  {Y.-J.}\ \bibnamefont {Kim}}, \ and\ \bibinfo {author} {\bibfnamefont
  {A.~D.}\ \bibnamefont {Christianson}},\ }\bibfield  {title} {\enquote
  {\bibinfo {title} {Spin-orbit coupled systems in the atomic limit: rhenates,
  osmates, iridates},}\ }\href {\doibase 10.1103/PhysRevB.97.235119} {\bibfield
   {journal} {\bibinfo  {journal} {Phys. Rev. B}\ }\textbf {\bibinfo {volume}
  {97}},\ \bibinfo {pages} {235119} (\bibinfo {year} {2018})}\BibitemShut
  {NoStop}%
\bibitem [{\citenamefont {Khomskii}\ and\ \citenamefont
  {Streltsov}(2021)}]{JH_3}%
  \BibitemOpen
  \bibfield  {author} {\bibinfo {author} {\bibfnamefont {D.~I.}\ \bibnamefont
  {Khomskii}}\ and\ \bibinfo {author} {\bibfnamefont {S.~V.}\ \bibnamefont
  {Streltsov}},\ }\bibfield  {title} {\enquote {\bibinfo {title} {Orbital
  effects in solids: Basics, recent progress, and opportunities},}\ }\href
  {\doibase 10.1021/acs.chemrev.0c00579} {\bibfield  {journal} {\bibinfo
  {journal} {Chem. Rev.}\ }\textbf {\bibinfo {volume} {121}},\ \bibinfo {pages}
  {2992--3030} (\bibinfo {year} {2021})}\BibitemShut {NoStop}%
\bibitem [{\citenamefont {Metropolis}\ and\ \citenamefont
  {Ulam}(1949)}]{Nicholas_1949}%
  \BibitemOpen
  \bibfield  {author} {\bibinfo {author} {\bibfnamefont {N.}~\bibnamefont
  {Metropolis}}\ and\ \bibinfo {author} {\bibfnamefont {S.}~\bibnamefont
  {Ulam}},\ }\bibfield  {title} {\enquote {\bibinfo {title} {The monte carlo
  method},}\ }\href
  {https://www.tandfonline.com/doi/abs/10.1080/01621459.1949.10483310}
  {\bibfield  {journal} {\bibinfo  {journal} {J. Am. Stat. Assoc.}\ }\textbf
  {\bibinfo {volume} {44}},\ \bibinfo {pages} {335} (\bibinfo {year}
  {1949})}\BibitemShut {NoStop}%
\bibitem [{\citenamefont {Ou}\ \emph {et~al.}(2014)\citenamefont {Ou},
  \citenamefont {Li}, \citenamefont {Fan}, \citenamefont {Wang},\ and\
  \citenamefont {Wu}}]{ou_2014}%
  \BibitemOpen
  \bibfield  {author} {\bibinfo {author} {\bibfnamefont {X.}~\bibnamefont
  {Ou}}, \bibinfo {author} {\bibfnamefont {Z.}~\bibnamefont {Li}}, \bibinfo
  {author} {\bibfnamefont {F.}~\bibnamefont {Fan}}, \bibinfo {author}
  {\bibfnamefont {H.}~\bibnamefont {Wang}}, \ and\ \bibinfo {author}
  {\bibfnamefont {H.}~\bibnamefont {Wu}},\ }\bibfield  {title} {\enquote
  {\bibinfo {title} {Long-range magnetic interaction and frustration in double
  perovskites
  $\mathrm{S}{\mathrm{r}}_{2}\mathrm{N}{\mathrm{i}}\mathrm{I}{\mathrm{r}}\mathrm{O}_{6}$
  and
  $\mathrm{S}{\mathrm{r}}_{2}\mathrm{Z}{\mathrm{n}}\mathrm{I}{\mathrm{r}}\mathrm{O}_{6}$},}\
  }\href {\doibase 10.1038/srep07542} {\bibfield  {journal} {\bibinfo
  {journal} {Sci. Rep.}\ }\textbf {\bibinfo {volume} {4}},\ \bibinfo {pages}
  {7542} (\bibinfo {year} {2014})}\BibitemShut {NoStop}%
\bibitem [{\citenamefont {Yang}\ \emph {et~al.}(2022)\citenamefont {Yang},
  \citenamefont {Xu}, \citenamefont {Lu}, \citenamefont {Zhou}, \citenamefont
  {Liu}, \citenamefont {Ma}, \citenamefont {Wang},\ and\ \citenamefont
  {Wu}}]{ke_2022}%
  \BibitemOpen
  \bibfield  {author} {\bibinfo {author} {\bibfnamefont {K.}~\bibnamefont
  {Yang}}, \bibinfo {author} {\bibfnamefont {W.}~\bibnamefont {Xu}}, \bibinfo
  {author} {\bibfnamefont {D.}~\bibnamefont {Lu}}, \bibinfo {author}
  {\bibfnamefont {Y.}~\bibnamefont {Zhou}}, \bibinfo {author} {\bibfnamefont
  {L.}~\bibnamefont {Liu}}, \bibinfo {author} {\bibfnamefont {Y.}~\bibnamefont
  {Ma}}, \bibinfo {author} {\bibfnamefont {G.}~\bibnamefont {Wang}}, \ and\
  \bibinfo {author} {\bibfnamefont {H.}~\bibnamefont {Wu}},\ }\bibfield
  {title} {\enquote {\bibinfo {title} {Magnetic frustration in the cubic double
  perovskite
  $\mathrm{B}{\mathrm{a}}_{2}\mathrm{N}{\mathrm{i}}\mathrm{I}{\mathrm{r}}\mathrm{O}_{6}$},}\
  }\href {\doibase 10.1103/PhysRevB.105.184413} {\bibfield  {journal} {\bibinfo
   {journal} {Phys. Rev. B}\ }\textbf {\bibinfo {volume} {105}},\ \bibinfo
  {pages} {184413} (\bibinfo {year} {2022})}\BibitemShut {NoStop}%
\bibitem [{\citenamefont {Hou}\ \emph {et~al.}(2015)\citenamefont {Hou},
  \citenamefont {Xiang},\ and\ \citenamefont {Gong}}]{hou_2015}%
  \BibitemOpen
  \bibfield  {author} {\bibinfo {author} {\bibfnamefont {Y.~S.}\ \bibnamefont
  {Hou}}, \bibinfo {author} {\bibfnamefont {H.~J.}\ \bibnamefont {Xiang}}, \
  and\ \bibinfo {author} {\bibfnamefont {X.~G.}\ \bibnamefont {Gong}},\
  }\bibfield  {title} {\enquote {\bibinfo {title} {Lattice-distortion induced
  magnetic transition from low-temperature antiferromagnetism to
  high-temperature ferrimagnetism in double perovskites
  ${A}_{2}\mathrm{F}{\mathrm{e}}\mathrm{O}{\mathrm{s}}\mathrm{O}_{6}$
  (${A}=\mathrm{C}{\mathrm{a}}, \mathrm{S}{\mathrm{r}}$)},}\ }\href {\doibase
  10.1038/srep13159} {\bibfield  {journal} {\bibinfo  {journal} {Sci. Rep.}\
  }\textbf {\bibinfo {volume} {5}},\ \bibinfo {pages} {13159} (\bibinfo {year}
  {2015})}\BibitemShut {NoStop}%
\bibitem [{\citenamefont {Kim}\ \emph {et~al.}(2008)\citenamefont {Kim},
  \citenamefont {Jin}, \citenamefont {Moon}, \citenamefont {Kim}, \citenamefont
  {Park}, \citenamefont {Leem}, \citenamefont {Yu}, \citenamefont {Noh},
  \citenamefont {Kim}, \citenamefont {Oh}, \citenamefont {Park}, \citenamefont
  {Durairaj}, \citenamefont {Cao},\ and\ \citenamefont
  {Rotenberg}}]{Kim_2008PRL}%
  \BibitemOpen
  \bibfield  {author} {\bibinfo {author} {\bibfnamefont {B.~J.}\ \bibnamefont
  {Kim}}, \bibinfo {author} {\bibfnamefont {Hosub}\ \bibnamefont {Jin}},
  \bibinfo {author} {\bibfnamefont {S.~J.}\ \bibnamefont {Moon}}, \bibinfo
  {author} {\bibfnamefont {J.-Y.}\ \bibnamefont {Kim}}, \bibinfo {author}
  {\bibfnamefont {B.-G.}\ \bibnamefont {Park}}, \bibinfo {author}
  {\bibfnamefont {C.~S.}\ \bibnamefont {Leem}}, \bibinfo {author}
  {\bibfnamefont {Jaejun}\ \bibnamefont {Yu}}, \bibinfo {author} {\bibfnamefont
  {T.~W.}\ \bibnamefont {Noh}}, \bibinfo {author} {\bibfnamefont
  {C.}~\bibnamefont {Kim}}, \bibinfo {author} {\bibfnamefont {S.-J.}\
  \bibnamefont {Oh}}, \bibinfo {author} {\bibfnamefont {J.-H.}\ \bibnamefont
  {Park}}, \bibinfo {author} {\bibfnamefont {V.}~\bibnamefont {Durairaj}},
  \bibinfo {author} {\bibfnamefont {G.}~\bibnamefont {Cao}}, \ and\ \bibinfo
  {author} {\bibfnamefont {E.}~\bibnamefont {Rotenberg}},\ }\bibfield  {title}
  {\enquote {\bibinfo {title} {Novel ${J}_{\mathrm{eff}}=1/2$ mott state
  induced by relativistic spin-orbit coupling in
  $\mathrm{S}{\mathrm{r}}_{2}\mathrm{I}{\mathrm{r}}\mathrm{O}_{4}$},}\ }\href
  {\doibase 10.1103/PhysRevLett.101.076402} {\bibfield  {journal} {\bibinfo
  {journal} {Phys. Rev. Lett.}\ }\textbf {\bibinfo {volume} {101}},\ \bibinfo
  {pages} {076402} (\bibinfo {year} {2008})}\BibitemShut {NoStop}%
\bibitem [{\citenamefont {Takagi}\ \emph {et~al.}(2019)\citenamefont {Takagi},
  \citenamefont {Takayama}, \citenamefont {Jackeli}, \citenamefont
  {Khaliullin},\ and\ \citenamefont {Nagler}}]{takagi_2019}%
  \BibitemOpen
  \bibfield  {author} {\bibinfo {author} {\bibfnamefont {H.}~\bibnamefont
  {Takagi}}, \bibinfo {author} {\bibfnamefont {T.}~\bibnamefont {Takayama}},
  \bibinfo {author} {\bibfnamefont {G.}~\bibnamefont {Jackeli}}, \bibinfo
  {author} {\bibfnamefont {G.}~\bibnamefont {Khaliullin}}, \ and\ \bibinfo
  {author} {\bibfnamefont {S.~E.}\ \bibnamefont {Nagler}},\ }\bibfield  {title}
  {\enquote {\bibinfo {title} {Concept and realization of {Kitaev} quantum spin
  liquids},}\ }\href {\doibase 10.1038/s42254-019-0038-2} {\bibfield  {journal}
  {\bibinfo  {journal} {Nat. Rev. Phys.}\ }\textbf {\bibinfo {volume} {1}},\
  \bibinfo {pages} {264} (\bibinfo {year} {2019})}\BibitemShut {NoStop}%
\bibitem [{\citenamefont {Clark}\ and\ \citenamefont
  {Abdeldaim}(2021)}]{clark_2021}%
  \BibitemOpen
  \bibfield  {author} {\bibinfo {author} {\bibfnamefont {L.}~\bibnamefont
  {Clark}}\ and\ \bibinfo {author} {\bibfnamefont {A.~H.}\ \bibnamefont
  {Abdeldaim}},\ }\bibfield  {title} {\enquote {\bibinfo {title} {Quantum spin
  liquids from a materials perspective},}\ }\href {\doibase
  10.1146/annurev-matsci-080819-011453} {\bibfield  {journal} {\bibinfo
  {journal} {Annu. Rev. Mater. Res.}\ }\textbf {\bibinfo {volume} {51}},\
  \bibinfo {pages} {495} (\bibinfo {year} {2021})}\BibitemShut {NoStop}%
\end{thebibliography}%

\end{document}